\documentclass[sigconf]{acmart}

\AtBeginDocument{%
  \providecommand\BibTeX{{%
    \normalfont B\kern-0.5em{\scshape i\kern-0.25em b}\kern-0.8em\TeX}}}

\copyrightyear{2023}
\acmYear{2023}
\setcopyright{rightsretained}
\acmConference[UIST '23]{The 36th Annual ACM Symposium on User Interface Software and Technology}{October 29-November 1, 2023}{San Francisco, CA, USA}
\acmBooktitle{The 36th Annual ACM Symposium on User Interface Software and Technology (UIST '23), October 29-November 1, 2023, San Francisco, CA, USA}
\acmDOI{10.1145/3586183.3606759}
\acmISBN{979-8-4007-0132-0/23/10}

\usepackage{color}
\usepackage{textcomp}
\usepackage{multirow}
\usepackage{subcaption}
\usepackage{bm}
\usepackage{wrapfig}
\usepackage{listings}
\newcommand{\new}[1]{\textcolor{black}{#1}}

\newcommand{\system}{\textsc{Synergi}}
\newcommand{\sys}{\textsc{Synergi }}
\newcommand{\systemws}{\textsc{Synergi }}

\newcommand{\txt}[1]{\texttt{#1}}

\newcommand*\cirnum[1]{\raisebox{.5pt}{\textcircled{\raisebox{-.9pt} {#1}}}}

\newcommand{\eg}{\textit{e.g., }}
\newcommand{\cf}{\textit{cf. }}

\newcommand{\ie}{\textit{i.e., }}

\newcommand{\mannw}[2]{\text{Mann-Whitney} $U$=#1, $p$=#2}
\newcommand{\wilcoxon}[2]{$\text{Wilcoxon}$ $W$=#1, $p$=#2}

\begin{document}

\title[\system]{\system: A Mixed-Initiative System for Scholarly Synthesis and Sensemaking}

\author{Hyeonsu B. Kang}
\email{hyeonsuk@cs.cmu.edu}
\orcid{0000-0002-1990-2050}
\affiliation{%
 \institution{Carnegie Mellon University}
 \streetaddress{5000 Forbes Ave}
 \city{Pittsburgh}
 \state{Pennsylvania}
 \postcode{15213}
 \country{USA}
}

\author{Sherry Tongshuang Wu}
\email{sherryw@andrew.cmu.edu}
\orcid{0000-0003-1630-0588}
\affiliation{%
 \institution{Carnegie Mellon University}
 \streetaddress{5000 Forbes Ave}
 \city{Pittsburgh}
 \state{Pennsylvania}
 \postcode{15213}
 \country{USA}
}

\author{Joseph Chee Chang}
\email{josephc@allenai.org}
\orcid{0000-0002-0798-4351}
\affiliation{%
  \institution{Allen Institute for AI}
  \streetaddress{2157 N Northlake Way \#110}
  \city{Seattle}
  \state{WA}
  \postcode{98103}
  \country{USA}
}

\author{Aniket Kittur}
\email{nkittur@cs.cmu.edu}
\orcid{0000-0003-4192-9302}
\affiliation{%
 \institution{Carnegie Mellon University}
 \streetaddress{5000 Forbes Ave}
 \city{Pittsburgh}
 \state{Pennsylvania}
 \postcode{15213}
 \country{USA}
}

\begin{abstract}
Efficiently reviewing scholarly literature and synthesizing prior art are crucial for scientific progress.
Yet, the growing scale \new{of publications} and \new{the burden of knowledge} make \new{synthesis of research threads more challenging than ever.}
\new{While significant research has been devoted to helping scholars interact with individual papers, building research threads scattered across multiple papers remains a challenge.}
\new{Most top-down synthesis (and LLMs) make it difficult to personalize and iterate on the output, while bottom-up synthesis is costly in time and effort.}
\new{Here, we explore a new design space of mixed-initiative workflows.}
\new{In doing so we develop a novel computational pipeline, \system, that ties together user input of relevant seed threads with citation graphs and LLMs, to expand and structure them, respectively.}
\new{\systemws allows scholars to start with an entire threads-and-subthreads structure generated from papers relevant to their interests, and to iterate and customize on it as they wish.}
In our evaluation, we find that \systemws helps scholars efficiently make sense of relevant threads, broaden their perspectives, and increases their curiosity.
We discuss future design implications for thread-based, mixed-initiative scholarly synthesis support tools.
\end{abstract}

\maketitle



\section{Introduction}
\begin{figure*}[t]
    \includegraphics[width=\textwidth]{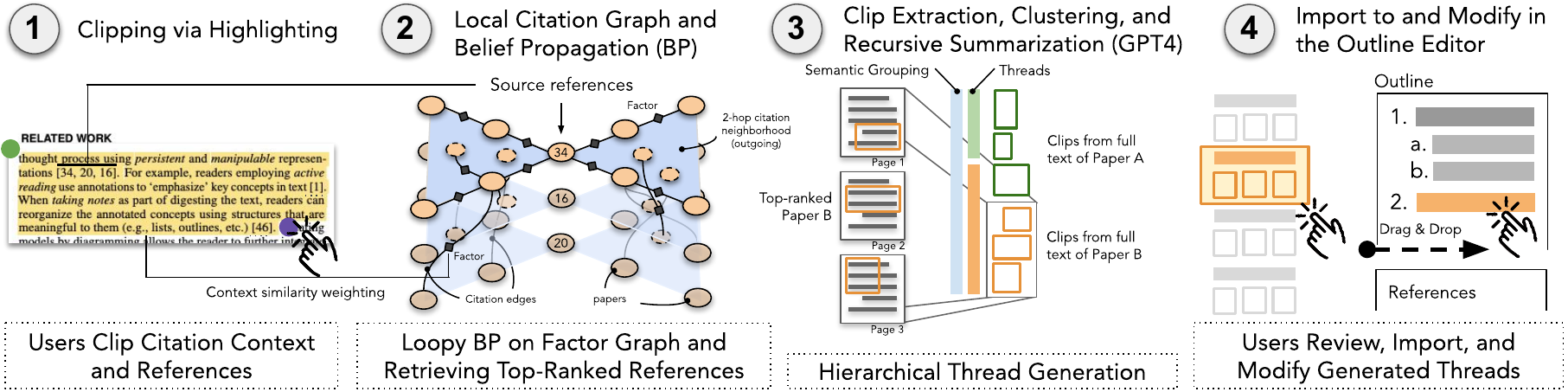}
    \vspace{-1.5em}
    \caption{Main stages of \system. (A) A scholar highlights a patch of text in a paper PDF that describes an interesting research problem with references. (B) The system retrieves important papers specifically relevant to the highlighted context in terms of how they have been previously cited by other scholars, via Loopy Belief Propagation over a local 2-hop citation graph from the seed references (Section~\ref{section:retrieval_algorithm}). (C) Relevant text snippets extracted from top-ranked papers are hierarchically structured and recursively summarized using GPT-4 in the chat interface (Section~\ref{section:gen_threads}). (D) The outline of threads, supporting citation contexts, and references are presented to the scholar for importing, modifying, and refactoring in the editor (Section~\ref{section:interaction_features} and \ref{section:outline_editor}).}
    \label{fig:main}
    \vspace{-1em}
\end{figure*}
Scientific and engineering innovations rely on synthesis of prior art: to know what approaches have been tried and identify most promising ideas for new problems; to unlock creative new ideas by combining existing ones; to reason about open challenges and unknown unknowns; and to contextualize one's research in a broader context of literature~\cite{knopf2006doing}. At the same time, scholarly synthesis is a cognitively difficult task because it involves many \new{demanding} inter-related steps in the process such as discovering relevant literature about a problem, reading and comprehending papers, collecting useful information and organizing it for further distillation, and recording and monitoring progress by developing an outline that summarizes current learning in the space~\cite{CiteSense}. Furthermore, scholarly synthesis becomes even more challenged by \new{expertise barriers for engaging with scientific literature due to deepening specialization}~\cite{bazerman1985physicists,hillesund2010digital,head2021augmenting}, the accelerating rate of growth~\cite{bornmann2015growth,jinha2010article}, and its increasingly interdisciplinary nature~\cite{okamura2019interdisciplinarity,van2015interdisciplinary}.

In order to synthesize knowledge scattered across multiple papers, scholars often employ iterative workflow\new{s} that involve multiple inter-related stages.
\new{Here, we focus on literature review workflows for high-level exploratory searches and synthesis in less familiar knowledge domains.}
\new{Such} workflow\new{s} can be characterized by their location on a spectrum of how much of the initiative is automated, between fully \textit{bottom-up} and fully \textit{top-down} workflows. 
Systems closer to the \textit{bottom-up} end of the spectrum such as Apolo~\cite{apolo} \new{and PaperQuest~\cite{paperquest}} allow users to explicitly save an interesting paper, \new{and Relatedly~\cite{relatedly} allows keyword queries for} expanding to additional papers and clips.
Threddy~\cite{threddy} \new{and Passages~\cite{han2022passages} enable highlighting of clips directly from documents for saving and organizing, allowing users to better maintain their context of reading in the process. However, in these workflows users are required to manually and repeatedly collect threads from documents whose cognitive and interaction costs can compound quickly.}

In contrast, systems near the \textit{top-down} end of the spectrum such as ConnectedPapers\footnote{\url{https://www.connectedpapers.com/}} and Metro Maps of Science~\cite{shahaf2012metro} provide \new{scholars} an initial visual overview of the research landscape to help \new{them} make sense of the structure of knowledge and discover interesting parts in it which can be especially useful for scholars new to a domain.
In addition, recent Large Language Models (LLMs)-based systems such as Galactica~\cite{taylor2022galactica}, ChatGPT\footnote{\url{https://chat.openai.com/chat}} and Google Bard\footnote{\url{https://bard.google.com/}} enable Q\&A-based interactions with knowledge domains which users can iteratively query.
However, the responses of such systems are similar to visual overviews described above in the sense that they are complete artifacts, rendering them less penetrable and useful for learning, iteration, and synthesis. \new{Although chat-based interfaces for LLMs can be helpful in various use case scenarios, they do not support users to easily extract useful parts of the output, to iterate on it by incorporating new information, or to incorporate supporting evidence, which are essential interactions for iteratively synthesizing knowledge from multiple documents.}
Furthermore, despite great potential for augmenting synthesis workflows, LLMs suffer from hallucination and falsehood (\cf~\cite{thorp2023chatgpt,bang2023multitask,black2022tweet}), rendering their outputs uncertain, less trustworthy, and needing manual inspection and verification.
\new{Instead, in this work we explore using LLMs as a component in a larger computational pipeline that constrains their scope to more tightly bounded summarization and synthesis goals, and enabling an alternative (non-chat) interactive interface that more directly supports users' needs.}

Here, we propose a novel mixed initiative workflow, \system, that augments scholars' existing synthesis workflows by providing them a structured outline view of research threads, which they can interactively review, curate, and modify.
\new{\systemws incrementally expands on user-curated threads that combine rich natural language descriptions and corresponding citations. Threads serve as boundary objects, translating user interests during literature exploration into signals for AI-based} outline \new{generation, supporting scholars to} move between the \textit{bottom-up} and \textit{top-down} workflows of scholarly synthesis, and help them combine the best of both worlds in the process.
\system-generated research threads relate specifically to a query clip and seed references, that may match only on a specific citation context within a paper rather than its entirety, and can directly help scholars with making sense of existing threads of research in an area and understanding their relations. \systemws accomplishes this by automatically retrieving a set of important papers from a 2-hop neighborhood on the citation graph and summarizing them in a hierarchical manner with a synthesized label for each parent node that captures the core commonality among its children. 
In contrast to prior approaches that supported largely manual \textit{bottom-up} synthesis workflows,
\systemws synthesizes threads from multiple papers and organizes them into a hierarchy that allows users to quickly discover most relevant threads and understand them through synthesis by other scholars, described in the citation contexts in their papers, that are provided together. Furthermore, in contrast to \textit{top-down} LLM-based workflows that may generate difficult-to-inspect black-box outputs, \system-generated threads maintain rich provenance and context to help users relate and inspect them further by following up on the source papers and the specific parts in their body text.

Through case studies and a controlled laboratory experiment where domain experts compared the quality of user-generated outlines from \systemws against those of a baseline system based on Threddy and a GPT4-based approach using the chat interface (henceforth referred to as Chat-GPT4) blind-to-condition, we found that \systemws resulted in the highest overall helpfulness ratings from expert judges.
Our quantitative analysis showed that the overall helpfulness of outlines from \systemws was \txt{1.6}-point higher compared to Chat-GPT4-generated outlines and \txt{2.6}-point higher compared to Threddy-based outlines (on a 7-point Likert scale).
In addition, experts judged that threads in the \systemws condition were better-supported with evidence from the literature compared to the Chat-GPT4 condition (\txt{+}$\Delta$\txt{3.3}) and the Threddy condition (\txt{+}$\Delta$\txt{2.3}; both on a 7-point Likert scale). Through quantitative and qualitative analyses of users' interaction logs, interviews, and responses to experience survey questions, we found that \systemws \new{encouraged} higher-level \new{thinking around} what existing salient threads of research are and how they divide the space, increased curiosity in them, and boosted confidence in conducting a literature review. We \new{also} found that these benefits likely came from efficiency gains over a \new{\textit{bottom-up}} Threddy-based baseline, and from gains in coverage of synthesis compared to a \new{\textit{top-down}} Chat-GPT4 baseline. We discuss these results and conclude with design implications for future AI-augmented scholarly synthesis systems and workflow designs.

In sum, the contributions of this paper include:
\begin{itemize}
    \item \system, a novel mixed-initiative workflow consisting of retrieval and organizational algorithms and interaction features to support scholarly synthesis.
    \item The results of a controlled laboratory and case studies involving expert judges and detailed quantitative and qualitative analyses of user interaction logs, interviews, and surveys uncovering the benefits and challenges of the approach.
    \item Implications for future workflow designs and relevant research inquiries in this area.
\end{itemize}

\section{Related Work}
\begin{figure}[h]
    \centering
    \includegraphics[width=\linewidth]{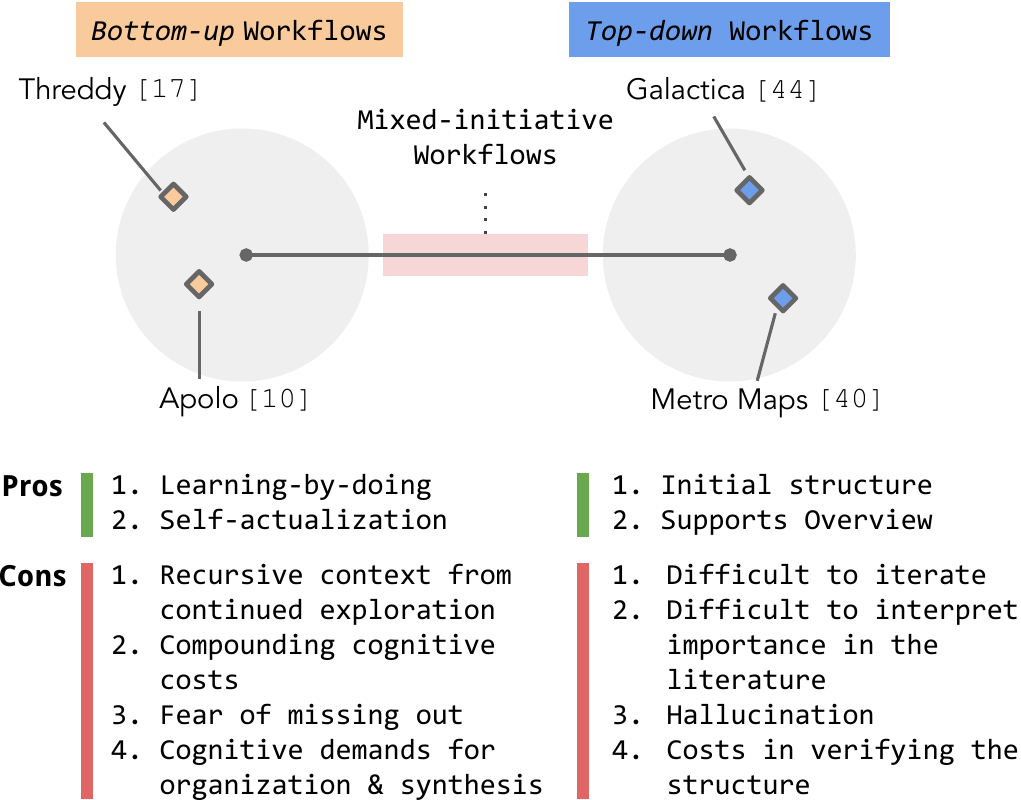}
    \vspace{-1em}
    \caption{The spectrum of workflows, with pros and cons in \textit{bottom-up} and \textit{top-down} workflows and relevant prior work.}
    \vspace{-1em}
    \label{fig:mixed_initiative_model}
\end{figure}
\subsection{\textit{Bottom-up} Scholarly Synthesis Workflows}

Workflows closer to the \textit{bottom-up} end of the initiative spectrum involve
practices such as forward and backward citation chasing and footnote chasing as integral parts for scholars traversing citation graphs to discover important papers related to a research problem~\cite{palmer2009scholarly}. However, these practices often suffer from fragmented information environments and piecemeal tooling~\cite{CiteSense} that add complexity to their workflows and \new{may} take focus away from synthesis. Moreover, while scholars can greatly benefit from reading the related work or introduction sections of a review paper that already synthesizes a relevant domain~\cite{teevan_2014}, they are not ideal for iterative exploration of the literature for synthesizing one's own review that may differ in focus, scope, or framing. 
Scholars would need to read multiple review papers for their synthesis~\cite{relatedly}, which can quickly \new{increase} the cognitive and interaction costs \new{(Fig.~\ref{fig:mixed_initiative_model} left, Con 1 \& 2)}.

Systems such as Apolo~\cite{apolo}, \new{CiteSee~\cite{chang2023citesee}} and Threddy~\cite{threddy} aim to \new{address these challenges by} helping scholars iteratively discover or group papers \new{(Fig.~\ref{fig:mixed_initiative_model} left, Con 3)}.
\new{Additionally, systems like Fuse~\cite{fuse}, ForSense~\cite{rachatasumrit2021forsense}, Passages~\cite{han2022passages}, Mesh~\cite{chang_mesh}, and Threddy~\cite{threddy}} support clipping and organizing \new{clips on-the-go, reducing the interaction and context-switching costs involved in the process}.
While helpful, interactions \new{in these systems focused on supporting users with manually saving clips or discovering additional papers, rather than synthesizing} knowledge after the early stages of discovery and foraging in sensemaking~\cite{pirolli2005sensemaking}.
\new{This leaves much of synthesis -- relating different clips, grouping references, and creating a coherent outline that describes multiple salient threads of research based on the data -- as manual work to scholars (Fig.~\ref{fig:mixed_initiative_model} left, Con 4).} 

\subsection{Systems that Support \textit{Top-down} Synthesis}
On the other end of the spectrum are systems such as ConnectedPapers\footnote{\url{https://www.connectedpapers.com/}} and Metro Maps of Science~\cite{shahaf2012metro} that provide a \textit{top-down} visual overview of the research landscape 
to aid scholars in making sense of the structure of knowledge space and to discover interesting parts in it. While such representations can serve as a great entry point to a knowledge domain that may be new to the user, they tend to not support additional user interactions beyond overview which limit their utility as a tool for synthesizing knowledge scattered across multiple papers \new{(Fig.~\ref{fig:mixed_initiative_model} right, Con 1)}.

Furthermore, recent systems' advances in Large Language Models (LLMs) such as Galactica~\cite{taylor2022galactica}, ChatGPT and Google Bard demonstrate impressive capabilities in answering user questions using the knowledge seemingly synthesized from the Web, and tools such as Ask Your PDF\footnote{\url{https://askyourpdf.com/}} show promising avenues for future systems that could support additional personalization and specification based on a set of user-curated documents. While these LLMs show great potential for augmenting scholarly synthesis workflows, they also suffer from challenges such as hallucination and falsehood (\cf~\cite{thorp2023chatgpt,bang2023multitask,black2022tweet}) that render their outputs uncertain, less trustworthy, and needing manual inspection and verification \new{(Fig.~\ref{fig:mixed_initiative_model} right, Con 3 \& 4)}. Moreover, the process of their computation is obscured~{\cite{arrieta2020explainable}} and less interpretable to users~\cite{linardatos2020explainable,zhang2021survey} which further reduces their chance of learning, iterating, and synthesizing based on their outputs \new{(Fig.~\ref{fig:mixed_initiative_model} right, Con 2)}.

\subsection{Systems that Augment Scholarly Discovery}
A large body of work exists in scholarly discovery~\new{\cite{lo2023semantic}}, including PaperQuest~\cite{paperquest} which allowed users to input query papers to receive other relevant papers based on citation relationships; Sturm~\cite{sturm2019design} which studied requirements for literature search systems and developed LitSonar where users could deploy nested queries to query over multiple sources of document streams; LitSense~\cite{litsense} which included multiple citation relation visualizations and supported filtering and querying for homing in on specific references for further exploration; search and recommender systems that leveraged citation graphs~\cite{feedlens,chi22_from_who_you_know,comlittee} to support relevance features in paper recommendations; \new{diversifying} scientific literature \new{search}~\cite{kang_augmenting_tochi,naacl2022_kang_augmenting}; and Relatedly~\cite{relatedly} that developed an approach to recommend relevant unexplored paragraphs in related work sections of papers.
Compared to prior work, \system's retrieval algorithm simultaneously optimizes for the \new{semantic} citation context similarity and the likelihood of a candidate paper building upon related threads of prior research.
\new{\systemws uses} Loopy Belief Propagation (LBP) \new{and} a novel message weighting scheme \new{on a local} citation graph \new{to} find papers most likely to be important in reviewing \new{related} literature, exploring new grounds beyond prior application of LBP in sensemaking over citation graphs~\cite{apolo}.

\section{Usage Scenario and Design Goals}
\subsection{\new{\systemws Usage} Scenario}
Consider a research scientist who wants to write a summary of notable threads of research \new{about \textit{how HCI professionals design human-centered AI systems}, a topic she recently} started exploring.
\new{She uses \systemws to open up the PDF of a paper that she saved earlier on the topic.}
As she reads through the introduction and related work sections of the paper, she finds several sentences with citations to prior work that describe notable advances in the literature.
\new{She clips the sentences by directly highlighting them in the PDF.}
\new{After saving a few clips from the paper, she is interested in a thread around \textit{challenges designers face with problem setting or ideation with AI}, and wonders what other related research threads there might be.}
\new{She quickly inputs her saved clips on \system.}

\new{Based on the input clips, \systemws recommends} threads and their high-level grouping that she can quickly scan to understand how different sub-group structure maps to the broader literature.
This understanding \new{allows her to} orient her attention towards specific areas that align with her interests.
\new{Because threads are organized in a hierarchy with rich provenance information such as the source references and exact excerpts from them that support each thread,} she can quickly identify threads that look particularly interesting in the literature, and find important references in them.
For some of the references she is not sure about or wants to understand in more details, she can \new{examine the} relevant sections in the paper \new{that support each thread that are presented together as excerpts}.

\new{After reviewing individual threads that were especially interesting to her, she can easily} curate \new{useful} threads, references, and contexts \new{from the provided hierarchy} into \new{an editor using drag-and-drop} where she \new{synthesizes} an outline and iterat\new{es} on it.
\new{After forming and iterating on her own initial synthesis outline, she has a few new references included in the outline that support individual threads.}
\new{She can quickly prioritize references that are more frequently cited in her hierarchy of research threads by using the group-by-reference view at the bottom of the editor.}
\new{This view gives a ranked list of references by the number and context of threads they appear under, which gives her an at-a-glance measure of `representativeness' a reference has to the threads included in the hierarchy.}
\new{She clicks on the first ranked reference to explore additional threads of research that may further expand the initial hierarchy she is building.}
\new{\systemws automatically opens the PDF in its reading interface, and she continues the literature review.}

\subsection{Design Goals}
Motivated by the challenges with existing tools and workflows described in the usage scenario, our design goals are as follows:
\begin{itemize}
\item[\textbf{[D1]}] When reading one research paper, allow scholars to clip passages and references of interests, and help them find important papers in the domain for synthesis, specific to query context and seed references.
\item[\textbf{[D2]}] Based on clips and references collected by a scholar, the system should provide a structured outline of salient research threads to support their synthesis across multiple papers.
\item[\textbf{[D3]}] Help scholars understand the specific research contexts described in each thread in detail, and verify their sources.
\item[\textbf{[D4]}] Help scholars review the system-generated threads, curate ones that most interest them into their own outline, and iteratively build upon it.
\end{itemize} 

\begin{figure*}[t]
    \centering
    \includegraphics[width=.9\textwidth]{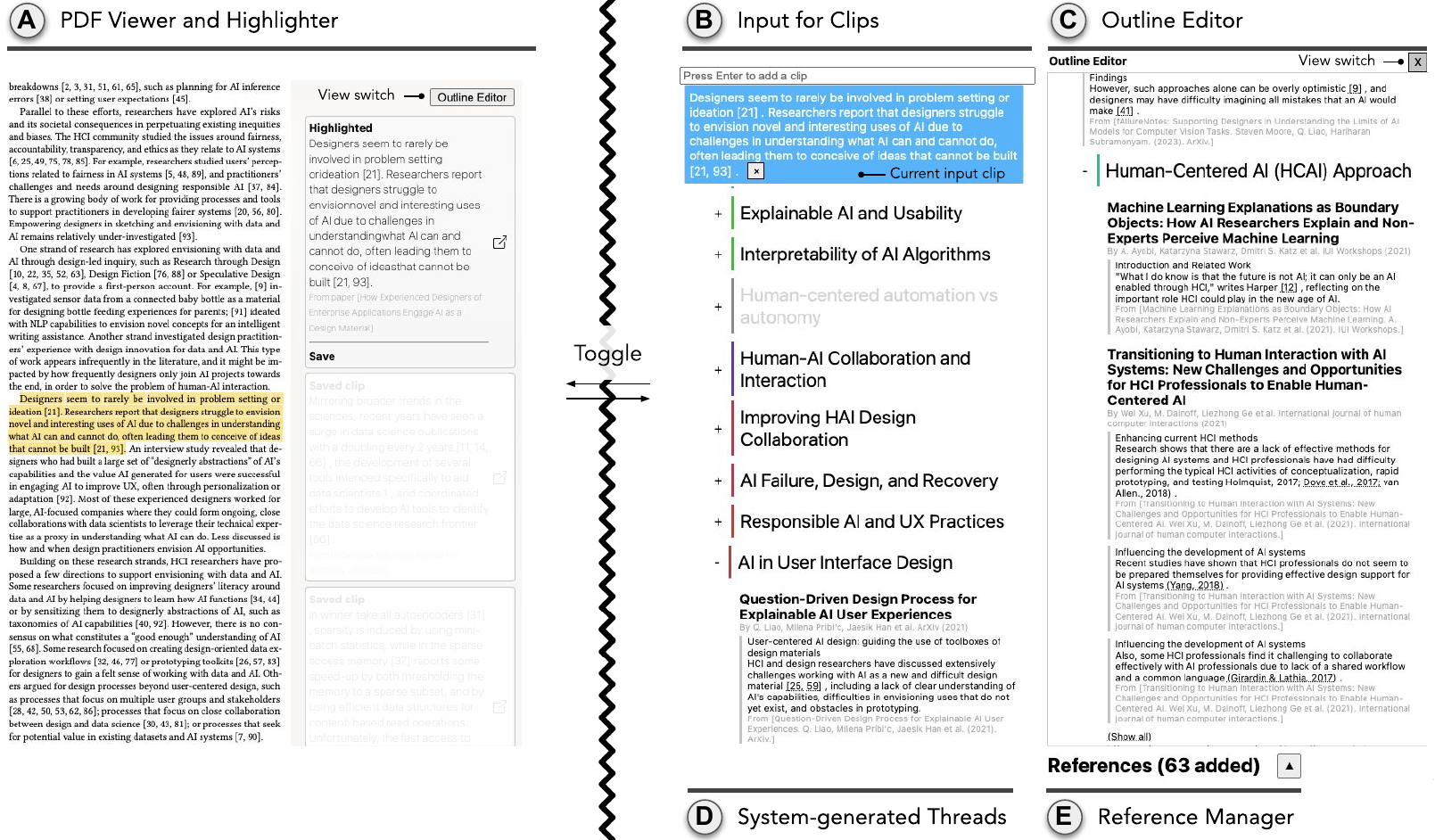}
    \vspace{-1em}
    \caption{Two main interfaces of \system. (A) The PDF viewer and in-text highlighter is similar to that of \textsc{Threddy}~\cite{threddy}, with a simplified stream of user-collected clips shown on the right. When the user clicks the `Outline Editor' button, the view switches to the editor mode. (B) In the top lefthand side corner is an input for user-collected clips where keywords of clips can be typed in to trigger a dropdown menu (not shown). Users can also click on ``Try these clips'' button to see the most recently saved clips for convenience of their reference (not shown). When the user adds a clip in the input, \systemws kickstarts the pipeline to generate a \txt{3}-level hierarch\new{y} of salient research threads in the literature specific to input clips. (C) Users can interact with the outline editor to curate interesting threads and citation contexts from the hierarchy (Section~\ref{section:outline_editor}). (D) \system-generated threads and grouped citation contexts are made draggable for user curation into the editor (Section~\ref{section:thread_recommendations}). (E) The reference manager automatically updates upon changes in the editor content (Section~\ref{section:citation_context_and_references}).}
    \label{fig:main_interface}
    \vspace{-1em}
\end{figure*}

\section{System Architecture}
The system consists of two primary backend algorithms and two sets of interface and interaction features corresponding to the design goals described above.
\subsection{Retrieving important papers specific to user's query citation context (D1)} \label{section:retrieval_algorithm}
\subsubsection{Background: Application of Loopy Belief Propagation for Sensemaking}
Loopy Belief Propagation (LBP)~\cite{bp} is a message-passing algorithm well-suited for iterative sensemaking over graphs that may contain cycles. LBP has previously been applied to sensemaking over citation graphs~\cite{apolo} due in part to its favorable qualities such as simultaneously being able to start from multiple entry points on a graph (e.g., multiple references in a user clipped paper passage), and supporting soft clustering (allowing each paper to belong to more than one research topics; see also Related Work in~\cite{apolo} for additional discussions of the algorithm's advantages over alternatives). While LBP on graphs with cycles may risk non-convergence, in practice the risk is extremely low on citation graphs due to the chronological ordering of citation edges leading to broken cycles and weak correlation~\cite{an2004characterizing}.

Different from Apolo~\cite{apolo}, in our workflow users start by specifying input that consists of \textit{the initial set of seed references} as possible exemplars on the citation graph, along with the \textit{citation context described in natural language in which they were referred to}. This setting does not assume user supervision is provided in an iterative manner throughout the process of discovery to prevent propagation of errors.

While previous use of Loopy BP over citation graphs only considered a set of user-provided seed papers to help discover additional papers~\cite{apolo}, users in \systemws clips passages and references as they read a paper to discover relevant research threads and papers. To incorporate this additional context (i.e., text passages) into Loopy BP,
we introduced a new multiplicative objective for context-sensitive message weighting (See Appendix~\ref{appendix:BP_details} for a detailed description), that goes beyond the constant message weighting scheme used in~\cite{apolo}. Intuitively, each component of the new multiplicative message weighting objective corresponds to the context similarity and reference overlap, respectively, optimization of which prioritizes papers that simultaneously meet the conditions of 1) that they are referred to in semantically related ways by other scholars in their literature reviews (typically appear in the introduction and related work sections of the paper) and 2) that they build upon related threads of research, represented by the overlapping set of references that they cited.
\new{Upon LBP convergence, probability distribution ranges from 0 to 1, with higher numbers representing greater relevance likelihood, generating rankings.}

\subsubsection{Construction of a factor graph using the 2-hop citation neighborhood}
We run the LBP algorithm over the local citation graphs sourced starting from the seed references provided in the user clip. In order to construct a candidate set of papers for retrieval (Fig.~\ref{fig:main}, \cirnum{2}), the system dynamically fetches the 2-hop citation neighborhood using each of the seed references in both directions (\ie incoming citations and references) using the Semantic Scholar APIs~\cite{kinney2023semantic}. For each seed paper referenced in a clip, this allowed \systemws to fetch up to \txt{50} most cited incoming or outgoing citations and \txt{50} references for each hop, resulting a total of \txt{50} * \txt{50} * \txt{2} = \txt{5,000} candidate papers. Once the 2-hop citation neighborhood is retrieved for each seed reference, we construct our factor graph with each unique candidate paper as a variable and use the citation edges as factors connecting the variables. To more deeply consider how each candidate paper is semantically relevant to the user clips, 
we also retrieve from the APIs information about each candidate papers including the titles and citing contexts. These information were stored as annotations on each edge in the factor graph. Since a paper can be cited by the same paper multiple times in different contexts, each edge may end up with multiple citation context annotations. Furthermore, each variable can be connected to multiple papers that have citation connections with it, allowing \systemws to capture different ways a candidate paper had been characterized by other scholars.

\subsubsection{Acquiring and parsing top-ranked paper PDFs} \label{section:system_pdf_acquisition}
Prior work~\cite{threddy} showed that specific citation contexts and synthesis already provided by other scholars (often appear in the related work or introduction sections of a paper) are useful for scholars' sensemaking and literature review. In order to extract them, we developed a full-text PDF acquisition and parse pipeline. First, we ran the LBP algorithm described above until convergence to find \txt{30} top-ranked papers to search for their full text PDFs. Then the pipeline initially searches the S2ORC corpus~\cite{S2ORC} to see whether a corresponding full text PDF URL is available for each paper.
In cases where a PDF URL was not available in the S2ORC corpus, the pipeline uses the Google Custom Search API\footnote{\url{https://developers.google.com/custom-search/}} to search for a matching paper title and its PDF URL using the ``filetype:PDF'' constraint. 
After obtaining a PDF file from the URL, the pipeline uses GROBID~\cite{grobid} to parse the PDF and extract the citation contexts along with metadata (\eg page number that the citation context appeared on, the header of the section containing the citation context, etc.) and the information of the references included in them to render in tooltips. Finally, if a candidate paper fails to fetch its PDF or be parsed, the pipeline defaulted to the paper title and abstract as its content.

\begin{figure*}[t]
    \centering
    \includegraphics[width=\textwidth]{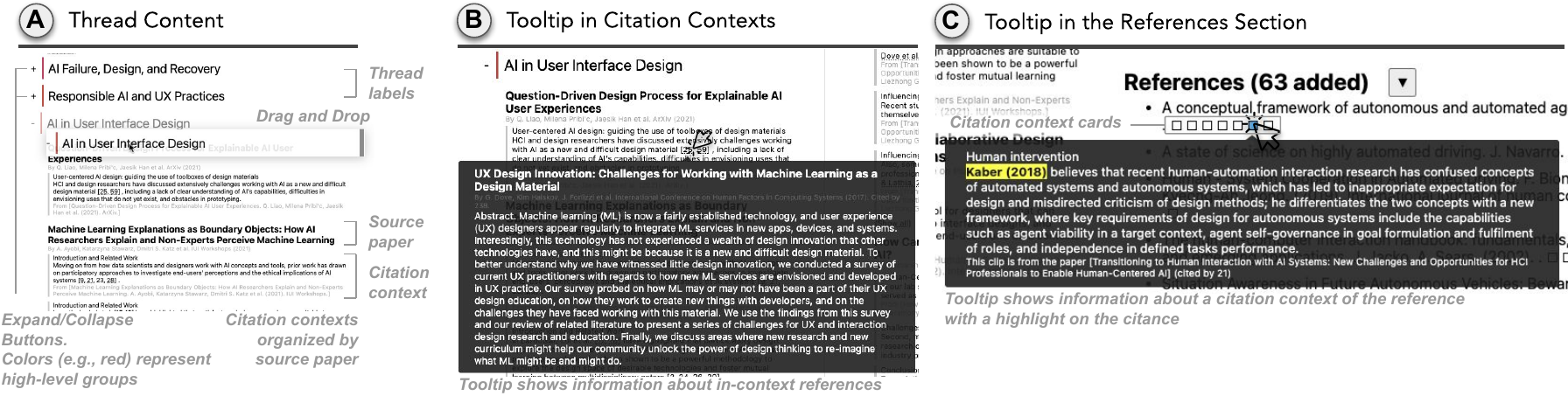}
    \vspace{-2em}
    \caption{Interface features. (A) The thread outline view is organized using an indented tree visualization. Threads and clips are visually differentiated using colors (the latter always featured a grey bar) as well as information organization. The citation contexts for each thread were grouped by the source papers and presented as a list. By default 3 contexts were shown; clicking on a [show more] button at the end of the list expands the list (not shown). (B) Mouse over each reference in a citation context (dotted and underlined for feature visibility) showed a tooltip that contained information about the reference. (C) The reference section at the bottom of the outline editor was automatically updated with each reference featuring citation cards; mouse over on a card showed a tooltip that contained the citation context information.}
    \vspace{-1em}
    \label{fig:interface_features}
\end{figure*}

\subsection{Generating Salient Threads of Research (D2)} \label{section:gen_threads}
Using the top-ranked papers from previous steps, \systemws generates a structured summary of multiple relevant threads of research in the area (Fig.~\ref{fig:main}, \cirnum{3}). This consisted of steps to home in on specific citation contexts in the papers, structure them into a hierarchy, and summarize them to capture core commonalities among the lower-level components in the hierarchy.

\subsubsection{Filtering citation contexts most relevant to seed clips}

To synthesize relevant information scattered across the multiple top-ranked papers identified from the retrieval algorithm (Section~\ref{section:retrieval_algorithm}) into a hierarchical structure using relevant text from them,
\systemws embedded the extracted citation contexts using \txt{text-ada-002}, and filtered those that have a higher average cosine similarity to seed clips than \txt{0.80}\footnote{determined through a small scaled experiment with five example clips during development}. \new{We used the S2ORC dataset~\cite{S2ORC} covering multiple citation intents for comparison. While not discerning context type, our pilot demonstrated functionality when combined with the context similarity thresholding}.

\subsubsection{Agglomerative clustering and tree-cutting} \label{section:agglomerative_clustering}
To present the most relevant topical clusters to the users, \systemws first uses the embeddings of the filtered citation contexts to measure how relevant they are to the user clip. For this, \systemws constructs a hierarchical structure from them using a unsupervised agglomerative clustering with the Ward linkage. We perform this using the \txt{fastcluster} package~\cite{mullner2013fastcluster}. Agglomerative clustering initializes citation contexts as singleton clusters and computes the ward distance of each pair to successively merge the most similar clusters. The result is a hierarchical binary tree (Fig.~\ref{fig:agglomerative_hierarchy}) where the height of the joint of branches represents the distance at which they were merged (the higher the height of the common ancestor of two leaf nodes on the hierarchy, the more distant they are as neighbors). The resulting binary tree is then converted into a \txt{3}-level hierarchy for the user to explore (see Appendix~\ref{appendix:tree_cutting} for a description of the rationale and the method).

\subsubsection{Recursively summarizing the children clusters} \label{section:recursive_summarization}
To help users explore the \txt{3}-level hierarchy, \systemws synthesizes labels for each parent thread that succinctly describes the underlying threads or citation contexts. In order to synthesize labels that are simultaneously coherent with the underlying children nodes' texts and are abstractions of them, we traverse the hierarchy in a bottom-up manner to recursively synthesize labels. We use Chat-GPT4 with a prompt (Fig.~\ref{fig:label_prompt} in Appendix~\ref{appendix:label_prompt}) that instructs it to summarize the underlying text using \txt{6} words or less. In each pass on a parent node, up to \txt{25} text snippets from its children were provided during prompting. Therefore, in the first pass the \txt{25} cluster citation contexts were added to the prompt and in successive runs, the text of the children clusters' synthesized labels were used. We also added a post-processing step to merge similar threads (see Appendix~\ref{appendix:post_processing_thread_merge} for a description of the rationale and the method) and assigned a unique color to each top-level thread such that the similarity among the children threads could be visually indicated later on the interface.

Finally, the \txt{3}-level tree structure with salient threads and their labels, along with the most relevant citation contexts attached to each, are returned to the front-end to render an overview of the relevant research landscape and salient threads in it.

\subsection{Interface Features (D2 \& D3)} \label{section:interaction_features}
\subsubsection{Walk-through of the interface}
Users on \systemws can highlight and clip relevant citation contexts directly from paper PDFs they are reading. Once they have one or more clips they are interested in investigating further, they switch to the editor view by clicking on the `Outline Editor' button from the PDF viewer (Fig.~\ref{fig:main_interface} \cirnum{A}). In the Outline Editor view, the user can select one or more from the list of saved clips to generate structured research threads related to the citation context and seed references included in selected clips (Fig.~\ref{fig:main_interface} \cirnum{B}).

Once the system finishes processing, the structured thread recommendations appear under the clip input (Fig.~\ref{fig:main_interface} \cirnum{D}). The user can review the content by scrolling through the list and by expanding/collapsing individual threads which contain the detailed information about citation contexts related to the thread, grouped by source papers. The colored bars on the left also provide users with high-level research areas to quickly orient themselves among the surfaced research areas and help guiding their attention to interesting ones. When the user identifies interesting threads, they can curate them into the outline they are building by dragging and dropping the threads from the list on the lefthand side into the outline editor (Fig.~\ref{fig:main_interface} \cirnum{C}), into the appropriate location on the hierarchy. The reference section below automatically updates based on the content changes in the editor, providing the users an easy access to information about papers that have been most cited across multiple threads and citation contexts, which help them prioritize what to read next. The user can continue the cycle by opening up a new paper in the PDF viewer and switching between outline editor. The user data persists for iterative development and refinement.

\subsubsection{Tree-structured thread recommendations} \label{section:thread_recommendations}
The tree-structured thread recommendations can be expanded and collapsed to reveal the relevant citation contexts below, which are grouped by source papers (Fig.~\ref{fig:interface_features} \cirnum{A}), to provide users with easy access to the source materials and increase the verifiability. Each thread label also featured a color bar on the left to indicate semantically similar groupings among different threads. Each citation context included the specific context found from the paper, the section header that it appeared in, as well as other metadata about the source paper.

\subsubsection{Citation context and reference tooltips} \label{section:citation_context_and_references}
To help scholars quickly gain additional information about the cited references in each citation context, each citation notation (\eg `[\underline{4}]') was rendered with a dotted underline (Fig.~\ref{fig:interface_features} \cirnum{B}), with an additional tooltip that reveals information about the reference such as its title, publication year and venue, number of citations, author names, and the abstract over a mouse-over. In the references section under the outline editor, each referenced paper was automatically updated when the content in the editor changes, and pulled in any citation contexts added in the editor that it was cited in. The grouped citation contexts were shown as squares next to the title (denoted as `citation context cards' in Fig.~\ref{fig:interface_features} \cirnum{C}), which revealed a tooltip that contains information about the citation context with the corresponding reference notation highlighted in the yellow over a mouse-over. 

\subsection{Drag-and-Drop Outline Editor (D4)} \label{section:outline_editor}
\begin{figure}[h]
    \centering
    \includegraphics[width=\linewidth]{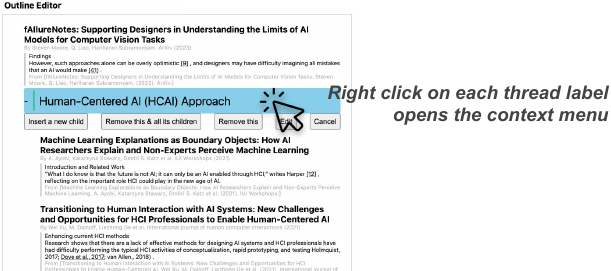}
    \vspace{-1.5em}
    \caption{Users could edit the outline either by adding a new thread or citation context into it using drag-and-drop, or by right clicking on each node in the editor.}
    \vspace{-1em}
    \label{fig:outline_menu}
\end{figure}
Threads or individual citation contexts were made draggable into the outline editor. Users could drop the dragged item into any thread node already in the editor or the default top-level thread (`Your Outline'). After the user drops an item to add to the editor, the references section below automatically updated to pull in any new references or new citation contexts for existing references (as shown in Fig.~\ref{fig:interface_features} \cirnum{C}). The added threads and citation contexts in the editor were interactive via right-clicking on them at which point the corresponding context menu was revealed. When a thread was right clicked, the following options were shown (Fig.~\ref{fig:outline_menu}):
\begin{itemize}
    \item[]\textbf{Insert a new child}: Add a new nested thread node.
    \item[]\textbf{Remove this \& all its children}: Completely remove the sub-tree rooted on this thread.
    \item[]\textbf{Remove this}: Remove only the clicked thread and moves all its children one level up (equivalent to merging).
    \item[]\textbf{Edit}: Edit the label of the thread.
    \item[]\textbf{Cancel}: Close the menu.
\end{itemize}
Right-clicks on citation contexts showed only the `Remove this', `Edit', `Cancel' options in the menu.

\section{Experimental Design} \label{sec:evaluation-design}
\begin{figure*}[t]
    \centering
    \includegraphics[height=1.3cm]{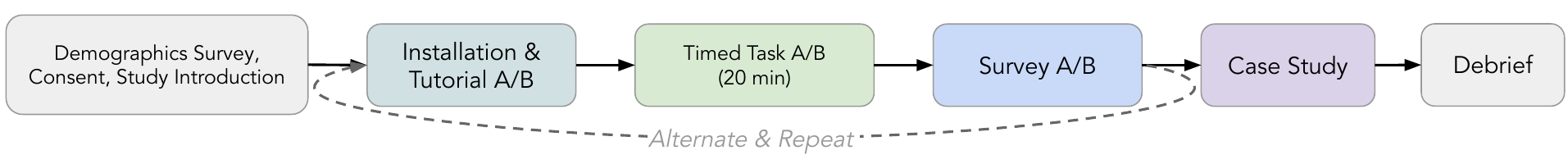}
    \vspace{-1em}
    \caption{The entire procedure of our study. The order of the middle section of the procedure was swapped based on the assignment (A/B). The order assignment was randomized and counterbalanced across participants (see text).}
    \label{fig:study_procedure}
\end{figure*}
\subsection{Objective \& Research Questions} 
Based on a user query as the input, 
we aimed to study how \system-generated threads of research and supporting clips can benefit scholars conducting literature review to cover the broader areas of research. 
We designed the timed tasks in the experiment to mimic the practice \new{of} coming up with a literature review outline for an assigned topic.
This is because scholars often craft intermediary outlines before arriving at a fully written article to structure their thoughts, synthesis, and exploration of the literature in earlier stages. We chose two different topics of research based on the papers that our expert judges were lead authors on~\cite{task_paper1,task_paper2}. To compare different conditions, we measure the quality of the outlines, the efficiency of constructing them, and the participants' perception of \system-generated threads and experience. We operationalized the quality of outlines as experts' judgment of the overall helpfulness, and thread-specific relevance, familiarity, and the goodness of the supporting citation context, on a Likert scale from 1 (Strongly disagree) to 7 (Strongly agree). We operationalized efficiency as the number of threads, clips, and references saved in the outline in a fixed amount of time, as well as the number of user actions taken to construct the outline. Our research questions were:
\begin{itemize}
    \item RQ1. Does \sys improve the quality of scholars' literature review outlines over the baselines?
    \item RQ2. Does \sys improve the efficiency of outline construction over the baseline?
    \item RQ3. What are perceived benefits and limitations of \system-augmented workflows?
\end{itemize}

\subsection{Participants} We recruited 12 \new{people} (10M/2F) for the study. \new{Participants'} mean age was 26.4 (SD: 2.11) and all actively conducted research at the time of the study (9 Ph.D. students and 3 Pre-doctoral Investigators). Participants' fields of studies included (multiple choices): HCI (10), NLP (4), Information Retrieval (1), Cognitive Science (1). We also recruited two experts (both female) to review participants' outlines. Both experts judges were 5th-year Ph.D. students with multiple first-authored and peer-reviewed publications in HCI venues. Their domains of research were `cross-functional AI teams in envisioning AI products and experiences' and `designing and building novel tools to help developers better annotate and share their learning materials'. The expert judges spent 1.5 hours to review participants' outlines and were compensated \$60 USD. The study lasted for 80 minutes and participants were compensated \$40 USD.

\subsection{Baseline Implementation}
\subsubsection{Baseline based on \textsc{Threddy}} \label{section:baseline_threddy}
The baseline system, based on prior work \textsc{Threddy}~\cite{threddy}, supported users in manually curating citation contexts via direct in-text highlighting in the PDF, with persisting clips across papers for increased context awareness, and featured a list of user curated clips on the lefthand side of the editor view that users could drag-and-drop into the editor easily. The user-curated clips replaced the system-generated outline provided in the treatment condition. All other interaction features \new{were identical}.
\begin{figure}[h]
    \centering
    \includegraphics[width=\linewidth]{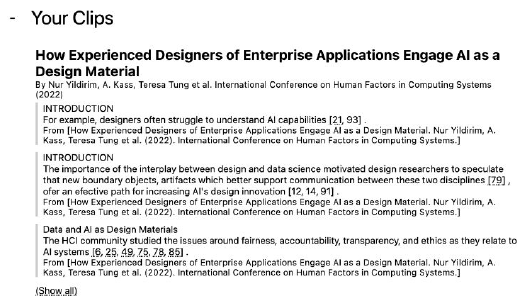}
    \vspace{-1.5em}
    \caption{The baseline system was based on \textsc{Threddy}~\cite{threddy} which supported clipping, persistence of clips across multiple papers, and an easy access to the outline editor where users could organize their own outlines using the self-curated clips.}
    \vspace{-1em}
    \label{fig:agglomerative_hierarchy}
\end{figure}
\subsubsection{Chat-GPT4 Baseline} \label{section:baseline_chatgpt4}
We also generated two literature review outlines for each paper \new{used in the main study with} Chat-GPT4 on OpenAI Playground\footnote{\url{https://platform.openai.com/playground}}. \new{Our prompts requested} \new{completion of} a literature review that \new{a scholar} has started, \new{given the} same citation context clip used in the treatment condition. \new{The prompts also included} a label of one \new{starter} thread. We replaced the citation notations \new{in the input context} with actual titles of the references, with clear demarcations, to provide further context about the research topic (see Fig.~\ref{appendix:baseline_prompt} in Appendix~\ref{appendix:baseline_prompt}). The temperature was set to \txt{1} with the maximum generation token length as \txt{2,048}. We repeatedly sampled two outlines for each of the two \new{input} paper-clip pair\new{s}. The \new{generated} outlines were then manually formatted/blinded (\eg removing auxiliary characters demarcating headers, reference notations, and unifying the style) for expert review.

\subsection{Procedure}
\new{Our main study simulates a literature review task using novel domains, with three conditions: a Threddy~\cite{threddy}-based condition (\S\ref{section:baseline_threddy}), the \systemws treatment, and a no-human Chat-GPT4 baseline (\S\ref{section:baseline_chatgpt4}). The study also triangulates the output quality through third-party expert assessment and case studies in familiar domains where participants can comment on the output quality. The inputs to the Chat-GPT4 baseline were prompts from Fig.~\ref{fig:baseline_prompt}.}
\subsubsection{Structure} We employed a within-subjects study \new{design} to compare \new{the outline construction process in two human-based conditions} -- \sys \new{and} Threddy-based \new{baseline}. \new{Chat-GPT4-generated outlines were blended to the outlines from the other two conditions for a blind expert quality assessment.} We chose two different research areas and topics for timed literature review tasks \new{with a randomized and counterbalanced presentation order}, and let individual participants choose personally interesting topic/paper for case studies in the end (\new{thus,} three tasks in total per participant). 
We randomly assigned systems to the topics for the timed tasks. We counterbalanced the order of presentation using 6 Latin Square blocks and randomized rows. Participants followed the following procedure in the study, which took place remotely using Zoom: Introduction, Consent, Demographics survey; Installation and Tutorial (detailed in Appendix~\ref{appendix:system_tutorials}) of the first system; Main task for the first system; Survey for the first system; Alternate and repeat for the second system; Case Study based on a personally interesting topic; Debrief. Participants were asked to share their screen during the timed tasks and think-aloud during the case studies.

\subsubsection{Timed Literature Review Tasks on Pre-defined Topics (20 mins each)} In each of the two timed tasks, participants were instructed to perform a literature review on a randomly assigned topic.
The interviewer provided the initial URL to the paper and pointed the participants to the exact location of the clip in each paper that contained the target problem statement. The scenario given to the participants was \new{framed} as `conducting a review of the relevant literature on behalf of their colleague, who is studying a related research question' \new{for motivating participants in unfamiliar domains.}

\subsubsection{Post-task Surveys} \label{subsection:surveys}
After each task, participants were administered a survey containing questions on their subjective feelings about the experience. Demand (both physical and cognitive) and overall performance were measured using the validated 6-item NASA-TLX scale~\cite{nasa_tlx}, where a more compact 7-point scale, mapped to the original 21-point scale, was instrumented~\cite{nasa_tlx_7point_use_case}. In order to probe the compatibility and adopt\new{a}bility of the technology with participants' existing literature review workflows, we included a modified Technology Acceptance Model survey from~\cite{tam_survey} (4 items). Furthermore, 8 types of benefits around discovery, sensemaking, outlining, curiosity, confidence, fear of missing out, and organization of clips and references were measured for each system (See Appendix~\ref{appendix:survey} for details of the questionnaire).

\subsubsection{Data Collection} We collected participant-generated literature review outlines at the end of each timed task. The outlines were then transformed into a spreadsheet while preserving the indentation of the original tree structure with additional columns on the left for experts' judgement. Each tree was traversed to tally the number of threads, clips, and references for each participant for analysis. During the experiment, participant's interaction traces (\ie timestamped action details during timed tasks) on each system were logged. The details of each timestamped action included a unique user ID, time of the action, the type of the action (\ie clip, import, create, move, edit, remove, merge), and corresponding details. Participants' think-alouds during the case study and debrief were recorded and transcribed.

\subsubsection{Experts' Evaluation} \label{subsection:expert_eval}
The participant-generated literature review outlines were anonymized and blended with two randomly sampled outlines from Chat-GPT4 for each paper (See Appendix~\ref{appendix:baseline_prompt} for the details of the prompts used). Therefore outlines were generated from three conditions in total, \textit{Baseline} -- the \textsc{Threddy}-based baseline system described in Section~\ref{section:baseline_threddy}, \textit{Treatment}, and the \textit{Chat-GPT4}-based baseline (Section~\ref{section:baseline_chatgpt4}). Experts reviewed each outline independently and blind-to-condition, and evaluated on the basis of the following 7-point Likert-scale (1: Strongly disagree, 7: Strongly agree) questions:
\begin{itemize}
    \item (Overall Outline Helpfulness) ``\textit{I found the outline with supporting context helpful for reviewing the relevant literature.}''
    \item (Thread Familiarity) ``\textit{I found the thread of research familiar.}''
    \item (Thread Relevance) ``\textit{I found the thread of research relevant.}''
    \item (Thread is Well-Supported by Citation Context) ``\textit{I found the thread to be well-supported by the specific citation context(s).}''
\end{itemize}
The overall helpfulness question was evaluated once per participant resulting in 12 data points in \textit{Baseline} and \textit{Treatment} conditions and 4 data points in the \textit{Chat-GPT4} condition; the three thread-level questions were evaluated once per thread per participant, leading to 108 data points (\ie 31 in \textit{Baseline}; 10 in \textit{Chat-GPT4}; and 67 in \textit{Treatment}) in total.

\subsubsection{Case Studies}
At the end of the timed tasks, the interviewer asked participants to find and open the PDF of a paper that they were personally interested in that was also in their domain of research using the treatment system. Each participant highlighted and clipped a patch of text (one sentence or longer) that described a particular research problem that also included at least one citation in it, then generated a list of threads using it in the same way as earlier in the timed task. Once the result has returned, the participants were asked to review the generated list of threads, their semantic grouping, the clips, and the references that the clips had originated from. The interviewer then asked questions a\new{bout} their quality, benefits, and limitations.

\subsubsection{Data Analysis} The mappings between the research questions and analyses of collected data are as follows.
\begin{itemize}
    \item RQ1. We analyzed the quality measures of the outlines, which were on a 7-point Likert scale, using non-parametric tests. For expert-evaluated overall helpfulness of outlines, the Wilcoxon's signed rank test was performed for the paired-samples data (\ie the Baseline vs. Treatment comparison) and the Mann-Whitney U test was performed for the independent data (\ie the Chat-GPT4 baseline vs. Treatment comparison). For independent data such as thread-level familiarity and relevance, the Mann-Whitney U test was used.
    \item RQ2. We analyzed the efficiency measures (\eg the average number of saved threads/clips/references in 20 minutes and the number of user actions taken to construct the outline) between the conditions using paired Student's t-test.
    \item RQ3. The Likert-scale and Likert-item responses in the survey data were analyzed using the non-parametric paired-samples Wilcoxon's signed rank test. Participants' comments during the case studies were transcribed and qualitatively analyzed using open coding. Participants' interaction logs were visualized as time graphs and used for triangulating relevant survey responses and qualitative data.
\end{itemize}

\section{Findings}
\begin{figure}[t]
    \centering
    \includegraphics[height=4.2cm]{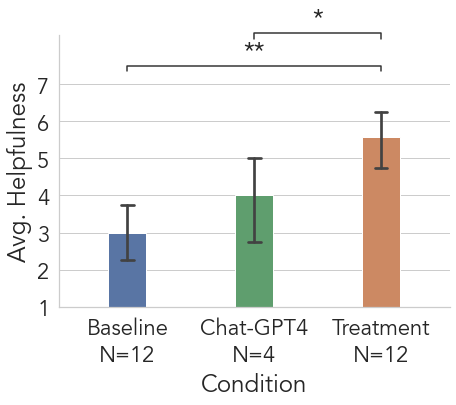}
    \vspace{-1em}
    \caption{The overall helpfulness judged by experts was the highest in Treatment (M=5.6), followed by Chat-GPT4 (M=4.0) and Baseline (M=3.0) conditions. The pairwise differences between Treatment and others were significant (see text).}
    \vspace{-1em}
    \label{fig:quality}
\end{figure}
\subsection{RQ1. Quality of Outlines}
\subsubsection{Higher quality outlines.} \new{U}sing \systemws, participants were able to generate literature review outlines that were rated as higher quality. The average expert judges' ratings on the overall helpfulness of literature review outlines in the Treatment condition was M=5.6 (SD=1.38), followed by the Chat-GPT4 condition (M=4.0, SD=1.41) and the baseline condition (M=3.0, SD=1.41) (Fig.~\ref{fig:quality}). Both differences between the Treatment and the Chat-GPT4 conditions (two-sided \mannw{7}{0.036}) and between the Treatment and the Baseline conditions (\wilcoxon{4}{0.003}) were significant. The experts were blind to the conditions that each of the outlines were generated under.

\subsubsection{Improved support while maintaining relevance and familiarity.}
\begin{figure*}[t]
    \includegraphics[width=.7\textwidth]{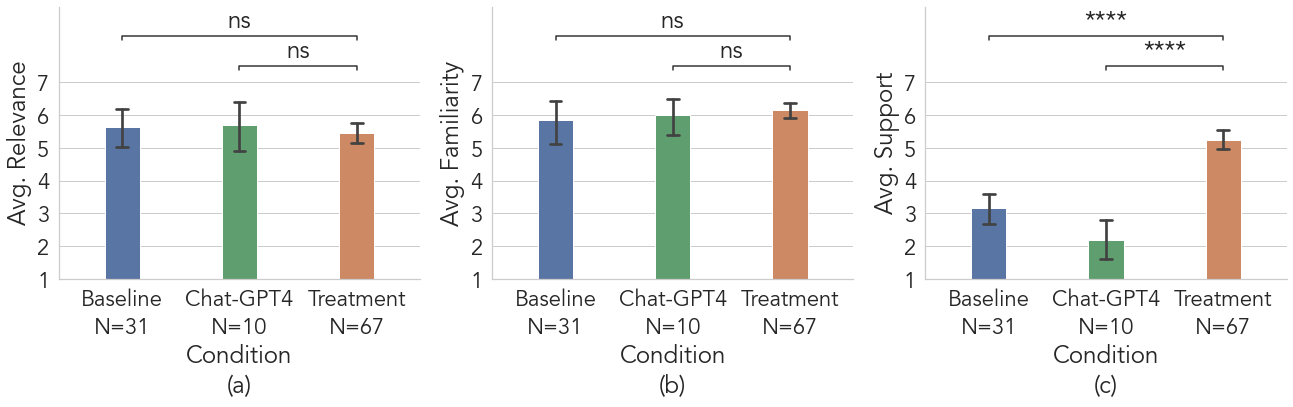}
    \vspace{-1.2em}
    \caption{Neither (a) average thread relevance nor (b) familiarity significantly differed between the conditions. (c) However, the average goodness of support from relevant citation context differed significantly, as it was judged higher in the Treatment condition (M=5.5, SD=1.25) than in the Chat-GPT4 (M=2.2, SD=1.03) or the Baseline (M=3.2, SD=1.27) conditions.}
    \vspace{-1em}
    \label{fig:thread_measures}
\end{figure*}
We further examined the overall outline helpfulness by comparing between the conditions their component threads' relevance, familiarity, and how well each thread was supported by relevant citation contexts found in the literature (Fig.~\ref{fig:thread_measures}). The results showed that the average thread relevance did not differ between the Treatment (M=5.4, SD=1.32) and the Chat-GPT4 (M=5.7, SD=1.34) conditions, nor between the Treatment and the Baseline (M=5.6, SD=1.74) conditions. Similarly, the average thread familiarity between the Treatment (M=6.1, SD=1.03) and the Chat-GPT4 (M=6.0, SD=0.94) conditions did not differ significantly, nor did the difference between the Treatment and the Baseline (M=5.8, SD=1.86) conditions. This suggests that while \systemws considered a large set of 2-hop references and citations (more than 5,000 candidate papers), it is able to maintain high relevance to the user query when presenting related research topics.

Further, the average support each thread received from relevant citation contexts differed significantly. Experts' judgement on the goodness of supporting citation contexts was the highest in the Treatment condition (M=5.5, SD=1.25) and positive (between `slight' (5) and `moderate' (6) levels), whereas in the Chat-GPT4 (M=2.2, SD=1.03; \new{$p<.0001$}) and the Baseline (M=3.2, SD=1.27; \new{$p<.0001$}\new{\footnote{\new{both were tested using two-sided Mann-Whitney}}}) conditions, it was negative and significantly lower. The goodness of support from relevant citation contexts also seemed to be a differentiating factor of the overall helpfulness of outlines among the conditions; while the relevance and familiarity measures for each thread were highly correlated (Kendall's $\tau$ \new{between .45 and .88, $p<.01$ in all cases}), support and other measures showed a weak relation \new{at best} (\new{relevance-support,} $\tau=0.21, p = .04$).

It is notable that despite the lack of supporting citation contexts, both the relevance and familiarity of an average thread generated by Chat-GPT4 tied with those of human-generated threads in the Baseline and Treatment conditions. However, our expert judges noted significant qualitative differences between the Chat-GPT4-generated threads from others, despite not knowing the sources of each outline during the evaluation. The judges proactively offered descriptions of how they differed qualitatively: ``[A Chat-GPT4-generated outline was] \textit{Probably the most coherent/thoughtful summarization and distillation of the source paper, but most of the stuff seems like something you could just get from reading only that paper and less of a literature review... no citations in any of the points... although the points are reasonable and feel like informed either by my work or other relevant source.}'' (E2); ``[After pointing out both Chat-GPT4-generated outlines] \textit{They seem like maybe someone read over some of the citations in my paper and pulled some points from that, but synthesis is generic. Overall, they are both not great as they don't include citations for the points outlined... Numbered lists in both outlines feel as if they were AI-generated, basically too generic to be useful without citations.}'' (E1).
\subsection{RQ2. Outline Construction Process}
\begin{figure*}[t]
    \centering
    \begin{subfigure}[t]{.4\linewidth}
        \includegraphics[height=3.7cm]{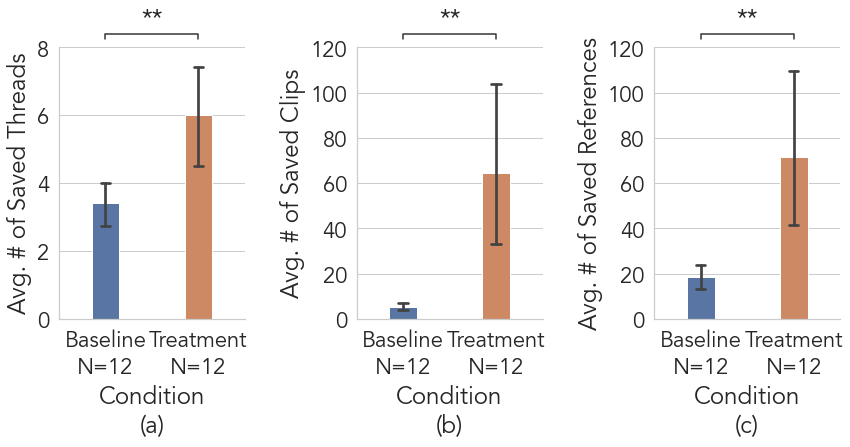}
    \end{subfigure}
    \quad
    \begin{subfigure}[t]{.4\linewidth}
        \includegraphics[height=3.7cm]{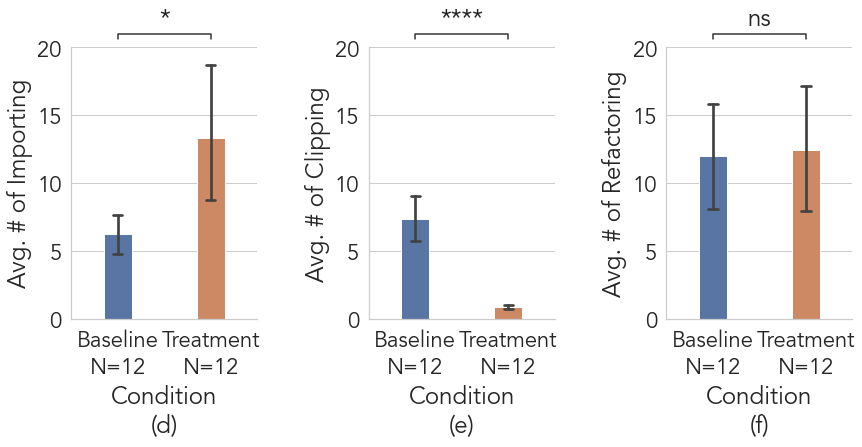}
    \end{subfigure}
    \vspace{-1em}
    \caption{The average (a) number of threads, (b) clips, and (c) references saved during the experiment (fixed length) were significantly higher in the Treatment condition than in the Baseline condition. (d) The differences in the saved numbers could be explained by how much more efficiently users in the Treatment condition imported system-generated outputs, rather than (e) spending time in manually clipping the relevant citation contexts, while (f) performing an overall similar amount of refactoring after adding new items to the outline editor.}
    \label{fig:construction}
\end{figure*}

\subsubsection{\systemws showed significant efficiency gains in the outline construction process}
\new{Comparing the outline construction process between the two human conditions\new{\footnote{\new{all comparisons in this section were performed using paired t-testing}}},} \new{t}he number of research threads, clips, and references saved in the duration of the experiment were all significantly higher in the Treatment than the Baseline condition (Fig.~\ref{fig:construction}a -- c). For threads, the average number saved was 6.0 (SD=2.76) in the Treatment condition vs. 3.4 (SD=1.16) in the Baseline condition (\new{$p=.01$}). The average number of saved clips was 64.3 (SD=66.27) in the Treatment condition vs. 5.5 (SD=2.81) in the Baseline condition ($p=.01$). The average number of saved references was also significantly higher in the Treatment (M=71.5, SD=63.40) vs. Baseline (M=18.4, SD=9.62) conditions (\new{$p=.01$}).

The higher numbers of saved items in the treatment condition could be explained by the overall higher frequency of `import' actions that users in the treatment condition performed (Fig.~\ref{fig:construction}d) compared to the baseline condition, instead of manually clipping (Fig.~\ref{fig:construction}e). On average, the users in the treatment condition performed 13.3 (SD=9.06) imports vs. 6.3 (SD=2.80) in the baseline (\new{$p=.02$}; Fig.~\ref{fig:construction}d) and 0.9 clipping (SD=0.29, Treatment) vs. 7.3 (SD=3.20, Baseline; Fig.~\ref{fig:construction}e) (\new{$p=.00002$}). The overall number of refactoring operations (\ie moving nodes in the outline editor, editing their labels, merging different thread nodes, removing nodes, creating a new parent thread) did not differ significantly between the two conditions (M=12.4, SD=8.44 in Treatment vs. M=12.0, SD=7.75 in Baseline; Fig.~\ref{fig:construction}f, \new{$p=.87$}), further suggesting that the efficiency gains originated from replacing the manual clipping of data with examining and importing the system-generated threads and clips in the treatment condition.

\subsubsection{\systemws supported both top-down and bottom-up workflows} \label{section:result_synthesis_workflows}
\begin{figure}[t]
    \centering
    \includegraphics[height=1.85cm]{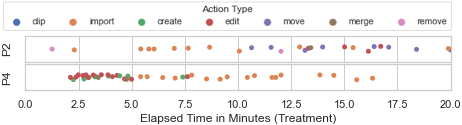}
    \vspace{-1em}
    \caption{(Top) A prototypical time-graph of user actions demonstrating a bottom-up approach of constructing the outline. (Bottom) Same for a top-down construction approach.}
    \vspace{-1em}
    \label{fig:prototype_actions}
\end{figure}
Interestingly, the users in the Treatment condition exhibited diverging patterns of constructing the outlines. Specifically, some users showed a pattern of top-down construction where they first carefully read through the problem statement and the rest of the source paper to come up with most salient threads of research in their mind before moving on to importing clips that fit those threads, and updating them when a new thread that expands or modifies the initial threads ideated by themselves. Fig.~\ref{fig:prototype_actions} (bottom) demonstrates a prototypical action time-graph which shows a densely populated area of refactoring in the beginning (\eg in the first 5 minutes in the graph) followed by successive importing. In contrast, Fig.~\ref{fig:prototype_actions} (top) demonstrates a prototypical time-graph for a bottom-up construction approach. In this case, the participant (P2) first imports a number of system-generated threads and clips onto the editor on the right, then moves on to refactor them (\eg past the 10 minute mark) to work towards a personally interesting outline. 

\subsection{RQ3. Perceived Benefits and Challenges with \system-augmented Workflows} \label{section:result_qualitative}
Quantitative analysis of survey results and qualitative analysis of interviews uncovered different types of benefits from \system, such as encouraging participants to gain a higher-level perspective about the literature, thinking about threads' relations, and increasing their curiosity. They also uncovered limitations of \system-augmented workflows such as additional refinement need related to identifying concepts at a similar level on the conceptual hierarchy, support for probing the relations among threads, and the desire to see explanatory relevance signals for user trust and acceptance.

\subsubsection{Reviewing \system-generated threads encouraged broader perspectives, sensemaking, and curiosity}
Participants commented on how having a list of automatically generated threads of research pushed them to think more broadly about the research space. P1 mentioned that the threads ``\textit{help you visualize the literature review outline in your head}'' and ``\textit{provide better and more context, especially useful for a new topic}'' (P1). Relatedly, P4 commented that:
\begin{quote}
    ``\textit{This is giving me a super-power to even begin to think at the level of `how are different threads of research dividing the space?', which would've been impossible for me to do otherwise.}'' -- P4
\end{quote}

Compared to how they typically conduct a literature in a new domain, they described feeling like saving a lot of time and cognitive effort (``\textit{I usually have to scroll back and forth so many times}'' -- P2; ``\textit{Overhead is significantly reduced... I can now just read, copy-paste, and re-organize stuff}'' -- P3) that would have otherwise interfered with forming higher-level perspectives. Participants' responses to the survey question: ``The system helped me discover relevant threads of research in the literature.'' also significantly favored the treatment condition (M=6.3, SD=0.75) over the baseline condition (M=3.3, SD=2.14; \new{$p=.009$}\new{\footnote{\new{participants' survey responses were compared using Wilcoxon's signed rank test}}}). Participants also felt as though the ``\textit{colors denoted good groupings of threads, for example this brown (color) shows a group about `Evaluation of toxicity' which was the core question in our research project.}'' (P7) and that ``\textit{the thread titles are pretty informative. I could easily tell what I should be paying attention to.}'' (P8). Interestingly, P1 commented on how ``\textit{it's refreshing to find threads on definitions and studies of `social capital' that may differ in non-western and global south's regional context of use}'' (P1) because manually chasing the citations alone tend to get you ``\textit{sucked into}'' the ``\textit{West-dominant}'' perspectives in the literature, since ``\textit{asymmetry in the citation behaviors exists between the western and non-western bodies of literature}'' -- P1.

Furthermore, participants' responses to survey questions: ``The system helped me make sense of relevant threads of research in the literature.'' (M=5.3, SD=1.66 in Treatment vs. M=4.3, SD=2.00 in Baseline, \new{$p=.088$}) and ``The system helped me outline a review of the literature.'' (M=6.1, SD=0.67 in Treatment vs. M=5.1, SD=2.02 in Baseline, \new{$p=.089$}) showed marginal significance between the two conditions at $\alpha=.10$.

Participants commented that the list of papers included in the references section of the outline, automatically extracted from the imported clips, was particularly relevant and contained ``\textit{inspiring papers to read in this area}'' (P7) and \new{some that one} participant wanted to take home (``\textit{Can I get a copy of the list on the left?}'' -- P11). P10 also described how the list ``\textit{Matches the threads and references that I curated for my own on-going literature review of the domain, which is good}'' (P10). Participants' responses to the survey questions also showed significant preference for the treatment condition over the baseline condition in terms of boosting their curiosity around different threads of research (M=6.0, SD=0.74 in Treatment vs. M=3.9, SD=1.73 in Baseline; \new{$p=.01$}), confidence in conducting the literature review (M=5.8, SD=0.94 in Treatment vs. M=4.0, SD=1.71 in Baseline; \new{$p=.01$}), and in reducing the fear of missing out on important research (M=5.2, SD=1.22 in Treatment vs. M=3.2, SD=1.64 in Baseline; \new{$p=.01$}) (See Appendix~\ref{appendix:survey} for the details of survey questions).

\subsubsection{Trade-offs between Completeness vs. Information Overload}
While participants reacted \new{favorably} towards the utility of \systemws in the context of the timed literature review outlining task (``\textit{This is a great starting point for a literature review}'' -- P10), they also commented on limitations that point to future \new{research} directions. One of the common concerns for longer-term use of \systemws raised by participants related to how to make sense of the quantity of threads presented to them. On the one hand, ``having this many, around 20 or so threads would overwhelm me easily'' (P10) and especially ``seeing similar threads, even though I like how they are grouped together using the same color, could really overwhelm me'' (P4). On the other end of the spectrum, seeing a widely varying number of threads returned for queries made P8 wonder if ``the result here is complete in this area because I only got 5 threads for this query. Or am I missing something important?'' (P8).

\subsubsection{Additional Support for Refining and Relating Threads}
Participants also commented on how in some cases the variations among the threads within the same high-level color group may be insignificant yet repeated, leading to visual clutter and information overload: ``\textit{[Newcomer Integration in OSS Projects] and [Newcomer barriers] are too similar, they can be merged}'' (P10); ``\textit{[Prompt engineering in NLP models] and [Prompting in Natural Langugage Processing] feel really similar}'' (P7). On the other hand, participants also pointed out threads that were seemingly too narrow in scope for them to be at the same level as other threads that seemed to synthesize across multiple papers: ``\textit{The [Skip-thought] thread is kind of weird to have be its own category because it's the name of a specific technique from a single paper.}'' (P6); ``\textit{[Numeric and logical reasoning] is focused on a very specific aspect of the papers in it, which I appreciate but feels too specific to be included in my review.}'' (P7).

P4 described how the threads of research helped him `lift' his perspective going into the literature review task which was beneficial. However, he also described how he was trying to interpret the relations and the order among different threads within each group and between differently colored high-level groups, and how he wished to ``\textit{also be able to reason about what the overlapping spaces are between the threads, for example in a `Venn diagram' of the research space... which is hard to do with a list of threads.}'' (P4).

An interesting sub-thread emerged in this topic when participants examined some of the `and' conjugated threads and found examples where the phrase before and after the `and' were at different levels of conceptual abstraction. Often the problematic cases featured one concept that felt too broad to be meaningful in relation to the other concept in the thread. For P10, a thread titled `[Augmenting scientific reading] and [machine learning]' was a clear demonstration of how the `and'-conjugated concepts could appear at different levels of abstraction, with the second concept in this specific example (\ie machine learning) being too high-level to be useful. Similarly P6 pointed out two examples, `[Text classification] and [feature weighting]' where the first concept was too broad to be meaningful, and `[Image Captioning] and [Computer Vision]' where the second concept ``\textit{did not feel like adding useful information}'' (P6).

\subsubsection{Scaffolding explanatory relevance information for trust and confidence in recommendations} Last but not least, participants wished to see additional information to understand how each thread was generated, and efficient at-a-glance information around which specific aspect in the query each clip is relevant to, in order to boost their confidence and trust in the recommendations. P10 said that:
\begin{quote}
    ``\textit{Understanding the sourcing mechanisms would help me gauge how much trust I should be lending to the system and stay vigilant for potential failure modes, because there are so many different kinds of relations that could be surfaced, for example `is it (relation) by authors? venues? publication years? topical similarity?' which makes me want to understand more.}'' -- P10
\end{quote}
For some, being able to group threads by a given paper was desired for helping orient their sensemaking process. P11 commented that ``\textit{In my process I move between papers when conducting a literature review... Here, some of the clips look similar to one another and I can see how the same paper is touching on different threads and I appreciate that the system has added clips from the same paper across multiple relevant papers... but it would be nice to be able to see which other threads that this paper has been added to so that I can quickly decide whether to read that paper in more details.}'' (P11). P12 commented that ``\textit{It would be helpful if I could see the connections between a thread and each clip in the thread because there are a lot of clips in this thread... and I want to quickly go through them, discarding the ones that look tangentially related.}'' (P12).

\section{Discussion}
\new{In this work, we design and develop \systemws as a mixed-initiative system for high-level literature exploration and scholarly synthesis, evaluate its benefits and challenges, and study implications for future systems aimed at augmenting scholars' workflows for synthesizing knowledge from many papers in a domain.}

\new{From evaluation studies we found that} study participants \new{engaged with} \system-generated threads of research to broaden their perspectives and free their cognitive bandwidth to focus more on higher-level thinking about salient threads and relations. 
\new{Interestingly, expert judges found} the Chat-GPT4-generated outlines \new{were} surprisingly well-synthesized \new{and} ``thoughtful,'' distill\new{ing} key points about the target problem statement.
\new{Supporting this observation,} the average expert-judged familiarity and relevance of threads did not differ between \new{both} human-generated and Chat-GPT4-generated threads.
\new{However, expert judges} also thought the helpfulness of outlines \new{depended significantly on the scope of its content and the quality of supporting citation context derived from relevant papers in the literature.}
By examining the outline construction process, we found that \new{the efficiency participants gained in foraging and making sense of research space in \system} allowed them to broaden their scope of synthesis and to incorporate more relevant papers and supporting citation contexts into their outline.
Taken together, these findings suggest that while LLMs such as GPT4 made remarkable advances in condensing scholarly text \new{on demand}, synthesis across multiple papers from the broader literature \new{with supporting context} remains a uniquely human capability today, albeit human scholars may be challenged by limited cognitive bandwidth while performing literature review and synthesis.

\subsection{\new{Thread-focused Workflows and Expansion}} \label{section:discussion_on_workflows}
Our examination of user interaction logs also revealed two salient behavioral patterns during synthesis around how and when they incorporated the \system-generated threads into their own outlines which we labeled as \textit{top-down} and \textit{bottom-up} synthesis workflows (\S\ref{section:result_synthesis_workflows}).
In the top-down workflow, users often started by processing the problem statement in more depth compared to the bottom-up workflow, and read surrounding contexts in the source paper \new{more deeply} to \new{form an initial understanding of their own, and then} distill their understanding into an initial outline.
In our evaluation participants using this workflow tended to have \new{deeper} prior knowledge in the research area that they could draw upon in creating the initial structure.
Once appropriate empty threads in their initial structure were identified, they subsequently imported relevant system-generated threads into them.

In contrast, in the bottom-up process participants often started off by iteratively importing system-generated threads into their editor on an individual thread basis, and creating ad-hoc parent threads when they find commonalities among existing threads.
Though lacking initial outline structures, this workflow was popular among the participants most of whom were new to the subject domains in the experiment.
\new{Bottom-up workflows on \systemws were made possible by using input threads as boundary objects for AI to pre-process other papers along related threads, forming an initial hierarchy with supporting references and context. Their popularity may also have been a result of increased user reliance, caused by \system's generation shifting the cost structure of sensemkaing~\cite{kittur_chi13_cost_benefit}, incentivizing user reliance for economic decisions~\cite{kool2018mental}.}

\new{Centering threads that capture core abstractions of references and citation contexts as first class objects in interaction design also opens up other new design spaces.}
\new{Possible future work includes thread-based AI-search and self-organization (\eg autonomously organizing snippets or pulling content from other papers to seamlessly expand the structure) or creativity-increasing retrieval (\eg targeting threads with generative potential, featuring core thread similarities and peripheral divergence).}
\new{Another future work direction could explore threads' different use contexts such as augmenting reading interfaces. For example, enabling an ambient `always on' mode that progressively suggests relevant snippets and threads from the same paper (\eg synthesized from later sections in the paper, allowing users to quickly scan the rest) or different papers (\eg supporting user transitions and further building on main threads).}

\subsection{Implications for Mixed-Initiative Workflows} \label{section:implications}
Our expert evaluation showed that fully AI-generated synthesis was competitive against outlines synthesized by human users in a manual or an AI-augmented workflow in \new{terms of} coherence and distillation \new{when the scope of synthesis was limited}.
Future LLMs with a sufficiently larger context window may overcome this issue via new capabilities in processing many papers at once.

However, even with an improved AI, a fully automated workflow may not be the \new{optimal} design for systems aimed at supporting scholarly synthesis.
`Putting in the work' during the literature review may be critical for scholars' learning and building up a necessary repository of knowledge \new{for successful synthesis later on}.
Rather than adopting a design that may disincentivize self learning and self-actualization~\cite{maslow1965self}, successful mixed-initiative systems therefore would need to consider tasks that AI augmentation can be most beneficial \textit{without} interfering with core cognitive tasks and human learning.
\new{For example, future workflow designs may} selectively delegate tasks involved in synthesis based on their high vs. low importance or the core vs. periphery division.
Scholars may specify a subset of research threads deemed peripheral to be further reviewed and summarized by an AI agent \new{taking an initial pass}, and \new{manually triage} whether newly identified threads from the summary merits \new{a closer look} from \new{them}, \new{minimizing sunk costs} in cases \new{of} irrelevant or uninteresting \new{results}.
\new{Another area is exploring designs to scaffold relevance signals for user comprehension of salient threads. Here, careful designs are needed for capturing users' interests at appropriate levels of thread abstraction initially, and supporting progressive refinement for iteration and better human-AI intent communication. As more relevance signals is not always better, the potential trade-offs between benefits and information overload must also be carefully examined.}

\subsection{\new{Beyond Chat-based Interfaces for LLMs}}
\new{Though helpful in various use scenarios, chat-based interfaces for LLMs significantly limit scholars' synthesis workflows. Such interfaces lack support for easy extraction of useful parts in the output and its iteration through incorporating new information or supporting evidence. Despite their lack of support, these interactions were common in study participants' workflows and were also regarded as adding significant value during their sensemaking and synthesis. Our expert evaluations confirmed that literature review outlines when generated on a chat-based interface had lower overall helpfulness ratings and included significantly less supporting evidence. Limited supporting evidence also had second-order downstream implications for reviewers' confidence in the output as well as scholars' further exploration and iteration.}

\new{Here, we present an alternative design approach that incorporates LLMs as part of a larger computational pipeline for interactive interfaces, that focused their processing to recursive summarization of relevant snippets from salient research articles on a topic. This approach enabled a mixed-initiative interface design where scholars could easily integrate parts -- useful threads -- from generated outputs and curate supporting evidence. Summarized threads benefited users by helping them support discover and prioritize new threads and references that they could explore further. Future interface designs may benefit from further exploring the design space of incorporating LLMs as components in computational pipelines, rather than standalone chat interfaces, yielding interaction designs that significantly benefit users in discovering, prioritizing, extracting, organizing, and synthesizing knowledge during sensemaking.}

\subsection{Limitations}
Though our evaluations uncovered new insights into scholarly synthesis workflows and implications for future mixed-initiative synthesis support tools, our experiments \new{were limited to} end-to-end evaluations of the pipeline.
\new{Additional} ablation studies \new{could} tease apart contributions from each component in the pipeline (\eg the \new{algorithms for} retrieval based on \new{novel} Loopy Belief Propagation; \new{for} formation of a thread-based hierarchy; and \new{for} recursive summarization using GPT4).
In addition, evaluati\new{ng} against a \new{future} baseline that has an expanded prompt context (\eg using multiple papers' text as input) \new{will contribute to whether} GPT4's synthesis capabilities generalize \new{to} multiple papers.
Furthermore, while our PDF acquisition and parsing was performant in the case studies \new{where} participants\new{'} personalized queries \new{were used}, scaling our approach to real-world scenarios with many users may require significant engineering \new{resources}.
A notable example here is how our system aimed to acquire and parse the full text PDFs for important papers, but it relied on best effort (by involving use of commercial APIs such as Google's Custom Search; \S\ref{section:system_pdf_acquisition}), without a guarantee of coverage.
While significant combined research and engineering efforts such as the S2ORC corpus~\cite{S2ORC} are notable in greatly increasing access to a large paper index with full text PDFs, we note that a significant portion of human knowledge \new{remains} locked in non-accessible PDFs, and concerted legal and institutional efforts may be required to make a significant step forward in this area. 

Finally, we believe that future empirical evaluations that go beyond the short duration for studies reported here, and in a more ecologically valid use context (\eg in a field deployment study) may uncover exciting new opportunities and challenges in this space.

\section{Conclusion}
In this paper we develop \system, a mixed-initiative system that supports scholarly synthesis and sensemaking of the scientific literature. In contrast to prior approaches that cater to either ends of the initiative spectrum (\ie \textit{bottom-up} or \textit{top-down} workflows), here we develop a novel approach to help scholars iteratively review the structure of literature related to a specific query context, curate important threads and references, and outline a useful review. Our evaluation that involved 12 participants and domain experts found that \systemws allowed users to create a higher-quality outline \new{for} a literature review, compared to a baseline based on a prior system, Threddy~\cite{threddy} and GPT4. We also found that \systemws achieves this through efficiency gains over the Threddy baseline. Moreover, we show that \systemws increased the coverage of synthesis while also enabling effective curation of supporting evidence from multiple papers over GPT4. Participants of the user studies found \systemws to be useful in broadening their perspectives about the literature, increasing curiosity while decreasing the fear of missing out on important research in the area. Finally, we conclude with implications for future mixed-initiative workflow designs for scholarly synthesis and interesting inquiries for research in the space. We believe more work is needed in this area to uncover new \new{mixed-initiative} workflow models and \new{to} envision improved systems that \new{can} help accelerate scientific innovation for all.


\begin{acks}
This work was supported by the Carnegie Mellon Center for Knowledge Acceleration, National Science Foundation (FW-HTF-RL, grant no. 1928631), the Allen Institute for Artificial Intelligence (Semantic Scholar), and the Office of Naval Research. We also thank the anonymous reviewers for their constructive feedback. In addition, we extend heartfelt thanks to our study participants, without whom this work would not have been possible.
\end{acks}

\bibliographystyle{ACM-Reference-Format}
\bibliography{zz_bib}

\appendix
\section{Detailed System Descriptions} 
\subsection{Loopy Belief Propagation Algorithm in \system} \label{appendix:BP_details}
\subsubsection{Background}
The use of LBP in prior work~\cite{apolo} was limited to a scalar conversion weighting of the probability (\txt{0.58}) when messages are exchanged between connected nodes in the graph. In other words, when the user assigns a category to a paper, the papers connected to that via citations would receive messages to increase their marginal probabilities of also being assigned the same category, regardless of the specific citation context. Furthermore, while this simple message weighting is a suitable configuration for interaction scenarios where the user provides iterative supervision over graph nodes (\ie user assigns a category $c \in C$ for each node $n$; each node state $s(n) \in \{c, \neg c, \text{not-seen}\}$), which can be used to correct subsequently propagating errors due to insensitivity to diverse citation relations, it is not suitable for our problem setting where no iterative supervision from the user can be supplied during the initial outline generation phase.

In contrast, in our problem setting the user input consists only of \textit{the initial set of seed references} as possible exemplars on the citation graph, along with the \textit{citation context described in natural language in which they were referred to}, without iterative supervision. 

\subsubsection{Running LBP with context-specific message scaling}
In order to prioritize papers that globally optimizes relevance and importance to the user input, we developed a multiplicative message weighting scheme which we assign to each factor in the factor graph to change the marginal probability after each local message passing between the two papers $v_i$ and $v_j$:
\begin{equation*}
    \frac{\left( \sum_{s \in S, k \in K}\text{sim}\left( \text{emb}(a_{i,j,k}), \text{emb}(c_s) \right) \right)}{\left| S \times K \right|} \times \frac{1}{1 + e^{-\left| \text{ref}(v_i) \cap \text{ref}(v_j) \right|}}
\end{equation*}
where $\{\forall s \in S: c_s\}$ is the set of seed clips, $\{\forall k \in K: a_k\}$ are the annotation texts stored on each edge between paper variable $v_i$ and $v_j$ (\ie note that $k \geq 1$ because the candidate paper's title text is always available even when no citation context text was found), $\text{sim}(\cdot, \cdot)$ represents the cosine similarity function that takes two embedding vectors as its input, $\text{emb}(\cdot)$ represents a text embedding using the Open AI's \txt{text-davinci-003} model, and $\text{ref}(\cdot)$ represents a function that takes a paper $v_i$ as its input to return the IDs of its referenced papers.

Intuitively, the first component of the multiplication corresponds to the average semantic similarity of possible pairings between the citation contexts in seed clips provided by the user and the citation contexts of the two papers. This is relevant because we are concerned with prioritizing papers with \textit{similarity specific to the query aspect}, rather than the entire paper's topical or thematic similarity to another paper. 

The second term of the multiplication corresponds to the degree of overlapping references between the two papers. Intuitively, the higher the number of overlapping references between the two papers, the more likely they would be building on similar threads of research, which can be a useful signal. Similar mechanism of triadic closure has been shown to be capable of surfacing missing friends~\cite{sharma_cosley_2013,abebe2022effect}, relevant paper recommendations~\cite{chi22_from_who_you_know}, and author recommendations~\cite{comlittee}. However, the effect of a small increase of the count of the overlapping references early on (\eg consider the effect from a step change $0 \mapsto 1$, in terms of the number of overlapping references between two papers; because there are many more papers that do not share any references, this step change may contain more discriminative information for classification than any other subsequent increases) may exhibit a steeper effect than the same difference at a higher base count of overlapping references. As such, we model the diminishing returns of this signal using the sigmoid function. Finally, the LBP is run until conversion\footnote{We did not encounter a non-converging case in the user studies.}.

\subsection{From Binary Tree to a \txt{3}-level Hierarchy} \label{appendix:tree_cutting}
\begin{figure}[h]
    \centering
    \vspace{-1em}
    \includegraphics[height=2cm]{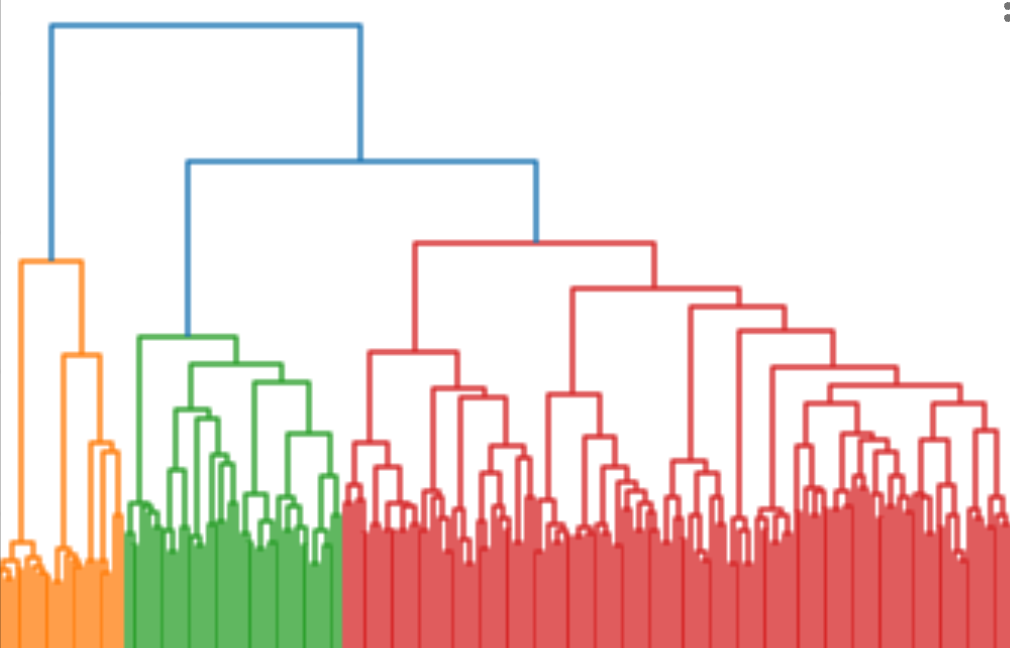}
    \vspace{-1em}
    \caption{Example hierarchy from agglomerative clustering.}
    \vspace{-1em}
    \label{fig:agglomerative_hierarchy}
\end{figure}
The resulting binary tree from the agglomerative clustering step in the algorithm (Section~\ref{section:agglomerative_clustering}) may contain within it the high-level hierarchy that resembles the structure that emerges from bottom-up coding of clips via this clustering process. However, in practice each thread in a literature review outline may have more than just two children citation contexts supporting it. For example, in the example binary tree outputted in Fig.~\ref{fig:agglomerative_hierarchy}, the tri-colored branches may correspond well to three distinctive research areas and thus need to be grouped into three semantic categories. Therefore, we condense and re-structure the binary tree in a way that hides the unnecessary complexity arising from the particular clustering method, while preserving the high-level semantic groupings  converted, into an \txt{3}-level N-ary tree by cutting it at \txt{3} different heights and pruning the branches that form elongated chains.

\subsection{Merging Similar Threads} \label{appendix:post_processing_thread_merge}
After piloting the synthesized labels of threads (Section~\ref{section:recursive_summarization}), we realized that the conversion of the full binary tree from agglomerative clustering into a \txt{3}-level hierarchy may have resulted in sub-groups that have similar citation contexts, that may be better described as a single larger high-level group. Therefore, we introduced a post-processing step that greedily merges parent threads that are highly similar in content from one another, thus reducing redundant sub-groups. We achieved this by using the pairwise cosine similarity of \txt{0.92} as threshold, which was determined from pilot testing. 

\subsection{Chat-GPT4 Prompt for Label Synthesis} \label{appendix:label_prompt}
\begin{figure*}[htbp]
\noindent\rule{\textwidth}{0.4pt}
\begin{lstlisting}
[System Message]
You are an agent that summarizes scientific articles.
- Follow the user's requirements carefully & to the letter.
\end{lstlisting}
\noindent\rule{\textwidth}{0.4pt}
\begin{lstlisting}
[User Message]
What is the topic commonly described in the following text snippets?
Summarize the topic succinctly (i.e., 6 words or less).
Reply with "Common topic: " followed by your response.
---
{input documents}
---
\end{lstlisting}
\noindent\rule{\textwidth}{0.4pt}
\vspace{-1.5em}
\caption{The prompt used to synthesize labels for each cluster using cluster members (\texttt{\{input documents\}}).}
\label{fig:label_prompt}
\end{figure*}
The input prompt to Chat-GPT4 consisted of a system message and a user message (Fig.~\ref{fig:label_prompt}). The outputs were generated using the OpenAI Playground interface\footnote{\url{https://platform.openai.com/playground}} in the chat mode using the GPT-4 model. The temperature was set to \txt{0}. The content of the user message was infilled with up to \txt{25} citation context text snippets in each cluster.

\section{Details of the study} 
\subsection{Tutorials} \label{appendix:system_tutorials}
Before participants start with each of the two main task with different conditions, they were given a tutorial of the assigned systems via screen sharing. The interviewer demonstrated a step-by-step installation process and the main features of each system using a prepared script that took around 10 minutes in each condition. In the baseline condition, participants were instructed to clip citances using in-text highlighter directly in the PDF, and switch between the editor and PDF viewer to organize saved clips into an outline. Participants could search for the PDFs of relevant papers on the Web using any popular search engines and continuously collect relevant clips from them. Participants in the treatment condition were instructed to start by reviewing the \system-generated threads and recommended clips to construct an outline.

\subsection{Chat-GPT4 Prompt for Literature Review} \label{appendix:baseline_prompt}
For the prompt in Fig.~\ref{fig:baseline_prompt}, the temperature was set to \txt{1} for repeated random sampling. The content of the user message was infilled using the content of each clip used in timed tasks, augmented by the titles of the references included in the clip.
\begin{figure*}[htbp]
\noindent\rule{\textwidth}{0.4pt}
\begin{lstlisting}
[System Message]
You are an assistant to a scientist who's conducting a literature review.
- Follow the user's requirements carefully & to the letter.
\end{lstlisting}
\noindent\rule{\textwidth}{0.4pt}
\begin{lstlisting}
[User Message]
Complete the following survey paper:

Title: Using Annotations for Sensemaking about Code - A Survey

### Code comments are not commonly used for keeping track of facts learned or open questions

Code comments are commonly utilized for keeping track of open tasks [START_REF]The emergent structure of development tasks.[END_REF][START_REF]Work Item Tagging: Communicating Concerns in Collaborative Software Development.[END_REF] and can be used as navigational aids [START_REF]How Software Developers Use Tagging to Support Reminding and Refinding.[END_REF][START_REF]Work Item Tagging: Communicating Concerns in Collaborative Software Development.[END_REF], but are not commonly used for keeping track of the other previously mentioned information needs developers have such as facts learned or open questions. This may be partially because the cost of externalizing this information, especially when the information may be incorrect, is too high [START_REF]Resumption strategies for interrupted programming tasks.[END_REF], and these code comments must then be cleaned up [START_REF]TODO or to bug.[END_REF].

###
\end{lstlisting}
\noindent\rule{\textwidth}{0.4pt}
\vspace{-1.5em}
\caption{The prompt used to generate outlines for expert review (showing content for one of the two papers used in timed tasks of the experiment). (Top) The system message component of the prompt. (Bottom) The user message component of the prompt. The temperature was set to \txt{1}. The prompt for the first paper in the timed task was similarly constructed, using the clipped citation context with demarcated (\eg enclosed within each \txt{[START\_REF]}...\txt{[END\_REF]} pair) reference titles.}
\label{fig:baseline_prompt}
\end{figure*}

\section{Detailed User Interaction Logs} \label{appendix:action_logs}
\begin{figure*}[t!]
    \centering
    \vspace{-1em}
    \includegraphics[width=\textwidth]{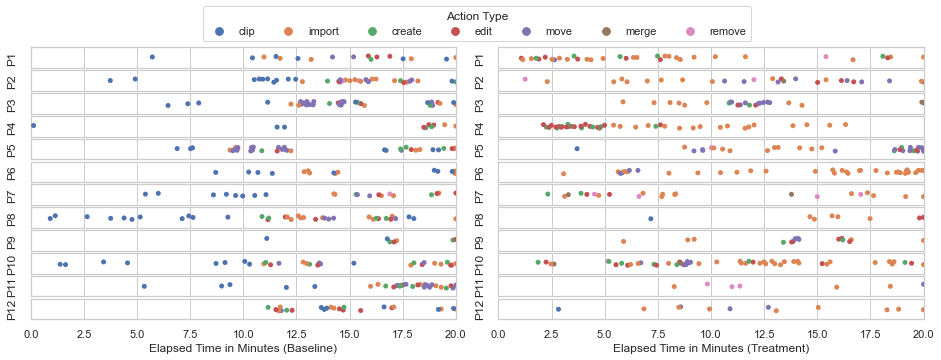}
    \vspace{-2em}
    \caption{User interaction logs on each system showing the timestamps of seven types of actions.}
    \label{fig:behavior_logs}
\end{figure*}

A time-graph of user actions in each condition is shown in Fig.~\ref{fig:behavior_logs}. 

\section{Full Survey Results} \label{appendix:survey}
Descriptions of survey items and participants' responses grouped by condition are presented in Table~\ref{table:full_survey}. Two-sided Wilcoxon's signed rank tests were performed to compute the $p$-values between conditions. See Section~\ref{section:result_qualitative} for discussions of the results.

\begin{table*}[t!]
    \centering
    \begin{tabular}{p{2.25cm} p{6.2cm} p{2.0cm} p{2.0cm} p{.85cm}}
    \toprule
    & \textbf{Description} & \textsc{Baseline} & \sys & \textit{p}-val. \\
    \midrule
    \multirow{4}{*}{{{1. NASA-TLX}}} & Sum of the participants' responses to the five NASA-TLX's~\cite{nasa_tlx} Likert-scale questionnaire items below. The original 21-point scale was mapped to a 7-point scale, similarly with~\cite{nasa_tlx_7point_use_case}. & \multirow{4}{*}{22.3 (SD=6.00)} & \multirow{4}{*}{17.9 (SD=4.19)} & \multirow{4}{*}{.08} \\
    \midrule
    1a. Mental & ``How mentally demanding was the task?'' & 4.8 (SD=1.36) & 4.3 (SD=1.42) & .34 \\
    \addlinespace[.1cm]
    1b. Physical & ``How physically demanding was the task?'' & 4.6 (SD=1.62) & 3.8 (SD=1.47) & .32 \\
    \addlinespace[.1cm]
    1c. Temporal & ``How hurried or rushed was the pace of the task?'' & 5.0 (SD=1.21) & 3.5 (SD=1.31) & $.003^{**}$\\
    \addlinespace[.1cm]
    \multirow{2}{*}{1d. Effort} & ``How hard did you have to work to accomplish your level of performance?'' & \multirow{2}{*}{4.4 (SD=1.44)} & \multirow{2}{*}{4.3 (SD=0.98)} & \multirow{2}{*}{.93} \\
    \addlinespace[.1cm]
    \multirow{2}{*}{1e. Frustration} & ``How insecure, discouraged, irritated, stressed, and annoyed were you?'' & \multirow{2}{*}{3.5 (SD=2.11)} & \multirow{2}{*}{2.0 (SD=1.21)} & \multirow{2}{*}{.08} \\
    \addlinespace[.1cm]
    \hline\hline
    \addlinespace[.1cm]
    \multirow{5}{*}{{{2. TAM}}} & Sum of the participants' responses to the 4 questionnaire items below adopted from~\cite{tam_survey} measuring the technological compatibility with participants' existing scholarly discovery workflows and the easiness of learning. & \multirow{5}{*}{19.1 (SD=4.48)} & \multirow{5}{*}{21.0 (SD=5.00)} & \multirow{5}{*}{.06} \\
    \midrule
    \multirow{4}{*}{2a. Compatibility} & ``\textit{Using the system is compatible with most aspects of how I search for scholars and their papers.}'' (The response Likert scales for this question and below are 1: \textit{Strongly disagree}, 7: \textit{Strongly agree}) & \multirow{4}{*}{4.1 (SD=1.51)} & \multirow{4}{*}{4.8 (SD=1.70)} & \multirow{4}{*}{.33} \\
    \addlinespace[.1cm]
    \multirow{2}{*}{2b. Fit} & ``\textit{The system fits well with the way I like to search for scholars and their papers.}'' & \multirow{2}{*}{4.7 (SD=1.83)} & \multirow{2}{*}{4.6 (SD=1.73)} & \multirow{2}{*}{.89} \\
    \addlinespace[.1cm]
    2c. Easy-to-Learn & ``\textit{I think learning to use the system is easy.}'' & 5.8 (SD=1.05) & 6.2 (SD=1.02) & .48 \\
    \addlinespace[.1cm]
    \multirow{2}{*}{2d. Adoption} & ``\textit{Given that I had access to the system, I predict that I would use it.}'' & \multirow{2}{*}{4.5 (SD=188)} & \multirow{2}{*}{5.4 (SD=1.73)} & \multirow{2}{*}{.15} \\
    \addlinespace[.1cm]
    \hline\hline
    \addlinespace[.1cm]
    \multirow{2}{*}{3. Discovery} & ``\textit{The system helped me discover relevant threads of research in the literature.}'' & \multirow{2}{*}{3.3 (SD=2.14)} & \multirow{2}{*}{6.3 (SD=0.75)} & \multirow{2}{*}{$.009^{**}$} \\
    \multirow{2}{*}{4. Sensemaking} & ``\textit{The system helped me make sense of relevant threads of research in the literature.}'' & \multirow{2}{*}{4.3 (SD=2.00)} & \multirow{2}{*}{5.3 (SD=1.66)} & \multirow{2}{*}{$.09$} \\
    \multirow{2}{*}{5. Outlining} & ``\textit{The system helped me outline a review of the literature.}'' & \multirow{2}{*}{5.1 (SD=2.02)} & \multirow{2}{*}{6.1 (SD=0.67)} & \multirow{2}{*}{$.09$} \\
    \multirow{2}{*}{6. Curiosity} & ``\textit{The system made me curious about different threads of research in the literature.}'' & \multirow{2}{*}{3.9 (SD=1.73)} & \multirow{2}{*}{6.0 (SD=0.74)} & \multirow{2}{*}{$.01^{*}$} \\
    \multirow{2}{*}{7. Confidence} & ``\textit{The system increased my confidence in reviewing the literature.}'' & \multirow{2}{*}{4.0 (SD=1.71)} & \multirow{2}{*}{5.8 (SD=0.94)} & \multirow{2}{*}{$.01^{*}$} \\
    8. Fear of Missing Out & ``\textit{The system reduced my fear of missing out on important research.}'' & \multirow{2}{*}{3.2 (SD=1.64)} & \multirow{2}{*}{5.2 (SD=1.22)} & \multirow{2}{*}{$.01^{*}$} \\
    {9. Organizing Clips} & ``\textit{The system helped me organize the clips I found.}'' & \multirow{2}{*}{5.7 (SD=1.15)} & \multirow{2}{*}{5.5 (SD=1.73)} & \multirow{2}{*}{$.79$} \\
    {10. Organizing References} & ``\textit{The system helped me organize the references I found.}'' & \multirow{2}{*}{5.2 (SD=1.59)} & \multirow{2}{*}{5.8 (SD=1.66)} & \multirow{2}{*}{$.34$} \\
    \addlinespace[.1cm]
    \bottomrule
    \end{tabular}
    \caption{Descriptions of full questionnaire items and responses grouped by condition. $p-$values are from two-sided paired samples Wilcoxon's signed rank tests.}
    \label{table:full_survey}
    \vspace{-2em}
\end{table*}

\end{document}